\documentclass[a4paper,12pt]{article}
\usepackage{amssymb, amsmath, mathtools}
\usepackage{epsfig}

\usepackage[pdftitle={Long Cycles in the Infinite-Range-Hopping Bose-Hubbard Models},pdfpagemode=UseOutlines,colorlinks=true,linkcolor=black,citecolor=black,pdfauthor={G. Boland},bookmarks=true,pdftex=true,unicode]{hyperref}

\makeatletter
\renewcommand{\section}{{\setcounter{equation}{0}}\@startsection%
{section}%
{1}%
{0mm}%
{-\baselineskip}%
{0.5\baselineskip}%
{\normalfont\large\bfseries}%
} \makeatother
\renewcommand{\theequation}{\arabic{section}.\arabic{equation}}

\makeatletter
\renewcommand{\subsection}{\@startsection%
{subsection}%
{2}%
{0mm}%
{-\baselineskip}%
{0.5\baselineskip}%
{\normalfont\normalsize\bfseries}}%

\newtheorem{proposition}{Proposition}[section]
\newtheorem{theorem}{Theorem}[section]
\newtheorem{lemma}{Lemma}[section]
\newtheorem{definition}{Definition}

\newenvironment{proof}{{\bf Proof:}}{\hfill$\square$\vskip.5cm}

\usepackage{color}

\def\*{{\phantom *}}

\newcommand{\sV}{{\hbox {\tiny{$V$}}}}
\newcommand{\sbeta}{{\hbox {\tiny{$\beta$}}}}

\newcommand{\Hone}{{\mathcal{H}_\sV}}

\newcommand{\trace}{\text{\rm{trace}}\,}

\newcommand{\Prob}{\mathbb{P}}
\newcommand{\Ex}{\mathbb{E}}

\newcommand{\astr}{^{\phantom *}}

\newcommand{\Ham}{H_\sV^{(n)}}
\newcommand{\cPart}{Z_\sbeta(n,V)}           

\newcommand{\cEx}{\Ex^n_\sV}   
\newcommand{\gEx}{\Ex^\mu_\sV}    
\newcommand{\gProb}{\Prob^\mu_\sV}    
\newcommand{\gHam}{H_\sV}
\newcommand{\gPart}{\Xi_\sV^\mu}           

\newcommand{\CC}{\mathbb{C}}
\newcommand{\RR}{\mathbb{R}}

\newcommand{\NN}{\mathbb{N}}

\newcommand{\FF}{\mathcal{F}(\Hone)}
\newcommand{\FFsym}{\mathcal{F}_{+}(\Hone)}

\newcommand{\la}{\langle}
\newcommand{\ra}{\rangle}

\newcommand{\Hcansym}[1]{{\mathcal{H}_{\sV,+}^{(#1)}}}
\newcommand{\Hcan}[1]{{\mathcal{H}_{\sV}^{(#1)}}}

\newcommand{\Happ}[1]{H_{#1,\sV}^{\textrm{\tiny{APP}}}}
\newcommand{\Hv}[1]{{H_{#1,\sV}}}

\newcommand{\qFFsym}{\mathcal{H}_{q, \sV}}

\newcommand{\Htilde}{\widetilde{H}}
\newcommand{\ii}{\mathbf{i}}
\newcommand{\jj}{\mathbf{j}}

\newcommand{\boldl}{\mathbf{l}}
\newcommand{\e}{\mathrm{e}}

\newcommand{\thermlim}{ \lim_{V \to \infty}}
\newcommand{\rholong}{{\rho^\mu_{{\rm long}}}}
\newcommand{\rhoshort}{{\rho^\mu_{{\rm short}}}}
\newcommand{\rhocond}{{\rho^\mu_{{\rm c\phantom{l}}}}}
\newcommand{\rhodens}{{\rho^\mu_{{\phantom{l}}}}}
\begin{document}


\phantom{.}
\vskip1.5cm
\begin{center}
{\Large Long Cycles
\vskip 0.1cm
in the Infinite-Range-Hopping Bose-Hubbard Model
}
\vskip 0.5cm
{\bf G. Boland}
\footnote{email: Gerry.Boland@ucd.ie}
\vskip 0.5cm
School of Mathematical Sciences
\linebreak
University College Dublin\\Belfield, Dublin 4, Ireland
\end{center}
\vskip 1cm
\begin{abstract}
\vskip -0.4truecm
\noindent
In this paper we study the relation between long cycles and Bose-Einstein condensation in 
the Infinite-Range Bose-Hubbard Model. We obtain an expression for the cycle density
involving the partition function for a Bose-Hubbard Hamiltonian
with a single-site correction.
Inspired by the Approximating
Hamiltonian method we conjecture a simplified expression for the short
cycle density as a ratio of
single-site partition functions. In the absence of condensation we prove that this simplification 
is exact and use it to show that in this case the long-cycle density vanishes.
In the presence of  condensation we can justify this simplification when 
a gauge-symmetry breaking term is introduced in the Hamiltonian. 
Assuming our conjecture is correct, we compare numerically the long-cycle density with the condensate
and find that though they coexist, in general they are not equal.

\end{abstract}
\newpage\setcounter{page}{1}
\section{Introduction}

Motivated by the path-integral formulation, in 1953 Feynman \cite{Fey1953} studied the relation between the statistical distribution of particles on permutation cycles and the occurrence of Bose-Einstein condensation (BEC).
He conjectured that the presence of long cycles is intrinsically connected to BEC.
Penrose and Onsager developed these arguments and observed that BEC should occur when the 
fraction of the total number of particles belonging to long cycles is strictly positive \cite{Pen.Onsa.}.
These concepts, which are now generally accepted, were made mathematically precise by 
S\"ut\H{o} \cite{Suto} who also proved the equivalence between the Bose-condensate density
and the density of the number of particles on long cycles in the case of the free and mean-field Bose gas
(see also Ueltschi \cite{Ueltschi2}).
Subsequently it was shown that this relation holds for the perturbed mean-field model of a Bose gas \cite{DMP}.
In our previous paper the validity of this hypothesis was tested in another model of a Bose gas, 
the Infinite-Range-Hopping Bose-Hubbard Model with Hard Cores \cite{BolandPule}. There it was shown that
while the existence of non-zero long cycle density and BEC coincide, these densities
were not necessarily equal.

This paper is the sequel to \cite{BolandPule}, where now the hard-core interaction is replaced
with a finite on-site repulsion to discourage but not forbid multiple particle occupation of individual sites.
The thermodynamics of this model have been studied by Bru and Dorlas \cite{BruDorlas}.
They show that the phase diagram for this model is much more complicated than in the hard-core case
as for low enough temperatures there are several critical values of chemical potential
which correspond to intervals of BEC (see Fig \ref{fig1}).
As in \cite{BolandPule}, we use standard properties of the decomposition of permutations into cycles,
to convert the grand-canonical sum into a sum on cycle lengths.
This makes it possible to decompose the total density $\rho=\rho_{\rm short}+\rho_{\rm long}$
into the density of particles belonging to cycles of finite length ($\rho_{\rm short}$)
and to infinitely long cycles ($\rho_{\rm long}$) in the thermodynamic limit.
We consider the relationship between Bose-condensation and long cycles in this model. 

The model is considered in the grand-canonical ensemble. In terms of the random walk 
representation, the particles hop from one site to another with probability depending on the occupation number of
the destination site -- the more particles on the site, the less likely another particle will hop there. 
Following \cite{BolandPule} we write the cycle density for the number of particles on a cycle of 
length $q$ in terms of a partition function of $q$ distinguishable particles interacting with the Boson system.
Again as in \cite{BolandPule} we exploit the fact that the hopping between the $q$ distinguishable
particles can be neglected, to obtain an expression for the cycle density in
terms of the ratio of the partition function for a Bose-Hubbard Hamiltonian
with a single-site correction and it without. Inspired by the Approximating
Hamiltonian method \cite{Bog4} we conjecture a simplified expression for the short
cycle density $\rho_{\rm short}$ as a ratio of
single-site partition functions. We prove that this simplification 
is exact in the absence of condensation and implies a zero long-cycle density in this case.
Unfortunately we are not able to prove this simplifying conjecture when BEC occurs, however we can go some way
towards justifying it by introducing a gauge-symmetry breaking term in the Hamiltonians. Assuming our conjecture is correct, we perform some simple numerical techniques to compare the long-cycle density with the condensate.
We find (as in \cite{BolandPule}) that though they coexist, in general they are not equal.

Before describing the layout of this paper, it is worth noting that BEC may be classified into three types 
(see \cite{vdB-L} and \cite{vdB-L-P}): type I/II when a 
finite/infinite number of one-particle quantum states are macroscopically occupied (resp.), and 
type III when no states are macroscopically occupied. The relation between the size of long cycles 
and these condensate types for the free Bose gas is considered in \cite{MBeau}.

The paper is structured as follows. In Section 2 we first describe the model and recall its thermodynamic properties
as stated by Bru and Dorlas \cite{BruDorlas}. In Section 3, by applying the general framework for 
cycle statistics described in \cite{DMP} (following \cite{Martin}), we form an expression for the density of 
cycles of length $q$ by isolating $q$ distinguishable particles from the boson 
field and show that we can neglect the hopping of these $q$ particles in the thermodynamic limit.
Section 4 deals with the above conjecture, proving its correctness in the absence of condensation 
and proving an equivalent result with the addition of a gauge-symmetry breaking term to the Hamiltonian
which is correct in the absence and presence of BEC.
Section 5 proves that in the absence of BEC the long cycle density is zero.

\section{The Model and Results}
The Bose-Hubbard Hamiltonian is given by
\begin{equation}								\label{B-H}
    H^{\mathrm{BH}}_\sV=J\!\!\!\!\!\!\sum_{x,y\in\Lambda_\sV\,:|x-y|=1}(a^\ast_x - a^\ast_y)(a\astr_x - a\astr_y)
    +\lambda\sum_{x\in \Lambda_{\sV}} n_x(n_x-1)
\end{equation}
where $\Lambda_\sV$ is a lattice of $V$ sites, $a^\ast_x$ and $a\astr_x$ are the Bose
creation and annihilation operators satisfying the usual commutation relations
$[a^\ast_x,a\astr_y]=\delta_{x,y}$ and $n_x=a^\ast_x a^{\phantom \ast}_x$.
The first term with $J>0$ is the kinetic energy operator and the
second term with $\lambda>0$ describes a repulsive interaction, as it discourages the presence of more
than one particle at each site. This model was originally introduced by
Fisher \textit{et al.} \cite{Fisher}.

The infinite-range hopping model is given by the Hamiltonian
\begin{equation}								\label{I-R}
	H_\sV
=	\frac{1}{2V}\!\!\!\sum_{x,y\in\Lambda_\sV}(a^\ast_x-a^\ast_y)(a\astr_x-a\astr_y)
	+\lambda\sum_{x\in \Lambda_\sV} n_x(n_x-1).
\end{equation}
This is in fact a mean-field version of (\ref{B-H}) but in terms of the kinetic energy rather
than the interaction. In particular, as with all mean-field models, the
lattice structure is irrelevant and there is no dependence
on dimensionality, so we can take $\Lambda_\sV=\{1,2,3, \ldots, V\}$.
The non-zero temperature properties of this model have
been studied by Bru and Dorlas \cite{BruDorlas} and by Adams and Dorlas \cite{AdamsDorlas}.
Also Dorlas, Pastur and Zagrebnov \cite{DPZ} considered the model in the presence of 
an additional random potential.

Bru and Dorlas applied the ``Approximating Hamiltonian'' method 
(see \cite{Bog4,Bog5}) to the 
Infinite-Range-Hopping Bose-Hubbard Model. In this method 
one performs the following substitution for the Laplacian term of the Hamiltonian:
\[
	\frac{1}{V} \sum_{x,y=1}^V a^\ast_x a\astr_y \rightarrow 
	\sum_{x=1}^V ( \bar{r} a\astr_x + r a^\ast_x ) - V|r|^2
\]
(some $r \in \CC$) to obtain the approximating Hamiltonian:
\begin{equation}								\label{Happrox}
	H^{\textrm{APP}}_\sV(r)
=	\sum_{x=1}^V n_x - 
	\sum_{x=1}^V ( \bar{r} a\astr_x + r a^\ast_x ) + V |r|^2
	+\lambda\sum_{x=1}^V n_x(n_x-1).
\end{equation}
Introduce a gauge breaking source $\nu \in \CC$ in both Hamiltonians (\ref{I-R}) and (\ref{Happrox}), by setting
$H_\sV(\nu) \vcentcolon= H_\sV - \sum_{x=1}^V (\bar{\nu} a\astr_x + \nu a^\ast_x)$ and 
$H^{\textrm{APP}}_\sV(r,\nu) \vcentcolon= H^{\text{APP}}_\sV(r) - \sum_{x=1}^V (\bar{\nu} a\astr_x + \nu a^\ast_x)$.
Then for all $\mu$ and $\nu$,
one finds that the pressures for these Hamiltonians are equivalent in the thermodynamic limit, i.e.
for large $V$ one obtains the estimate
\[
	0 \le p_\sV[H_\sV(\nu)] 
	- \sup_{r\in \CC} p_\sV[H^{\textrm{APP}}_\sV(r,\nu)] \le O(V^{-1/2})
\]
where for a Hamiltonian $H$, $p_\sV[ H ] \vcentcolon= p_\sV[ H ](\beta, \mu)$ denotes the corresponding
grand-canonical pressure. Henceforth 
the $\beta$ and $\mu$ dependencies are assumed unless explicitly given.

With this technique, Bru and Dorlas managed to obtain the limiting pressure
and showed that in some regimes Bose-Einstein condensation occurs. 
They proved the following result:

\begin{theorem}								\label{thm_soln}
The pressure in the thermodynamic limit for the Infinite-Range-Hopping Bose-Hubbard Model,
$p(\beta, \mu) \vcentcolon= \thermlim p_\sV [H_\sV]$, is given by
\begin{equation}								\label{pressure}
	p(\beta, \mu) = \sup_{r \ge 0} \bigg\{  -r^2 + \frac{1}{\beta} \ln \trace_{\mathcal{F}_+(\CC)} 
	\e^{-\beta h(r) } \bigg\}
\end{equation}
where 
\[
	h(r) \vcentcolon= (1 - \mu)n + \lambda n(n-1) - r ( a + a^\ast)
\]
is a single site Hamiltonian with creation and annihilation operators $a^\ast$ and $a$,
and with number operator $n=a^\ast a$. Note that it is sufficient to take the supremum over the set of
non-negative real numbers.

The Euler-Lagrange equation for the variational principle is
\begin{equation}								\label{euler-lagrange}
    2r
=  \big\la a + a^\ast \big\ra_{h(r)}
\vcentcolon=  \frac{\trace_{\mathcal{F}(\CC)} (a+a^\ast) \e^{-\beta h(r)}}
      {\trace_{\mathcal{F}(\CC)} \e^{-\beta h(r)}}.
\end{equation}
Moreover the density of the condensate is exactly given by
\[
	\rhocond \vcentcolon= \thermlim \frac{1}{V^2} \sum_{x,y=1}^V \big\la a^\ast_x a\astr_y \big\ra_{H_\sV} = r_\mu^2.
\]
where $r_\mu$ is the largest solution of $(\ref{euler-lagrange})$.
\end{theorem}

\begin{figure}[hbt]
  \begin{center}
      \includegraphics[width=16cm]{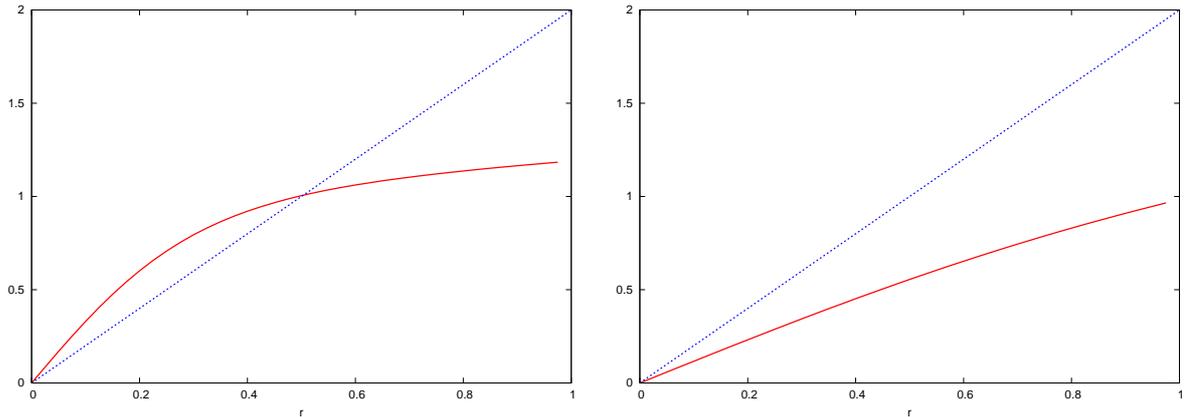}
  \end{center}
  \caption{\it Comparison of $2r$ with $\la a+a^\ast\ra_{h(r)}$ with $\beta=4$, $\lambda=5$ %
    for the cases $\mu=1.5$ (condensation) and $\mu=5$ (no condensation).}
  \label{figs2}
\end{figure}

Equation (\ref{euler-lagrange}) can have at most two solutions. 
Clearly $r=0$ is always a solution. When $\beta$ is large enough, for certain values
of $\mu$ a second non-zero solution may appear (see Fig \ref{figs2}). So $r_\mu \vcentcolon= 0$
unless a second solution $r>0$ exists, in which case $r_\mu \vcentcolon= r$. 

\begin{figure}[hbt]
  \begin{center}
      \includegraphics[width=16cm]{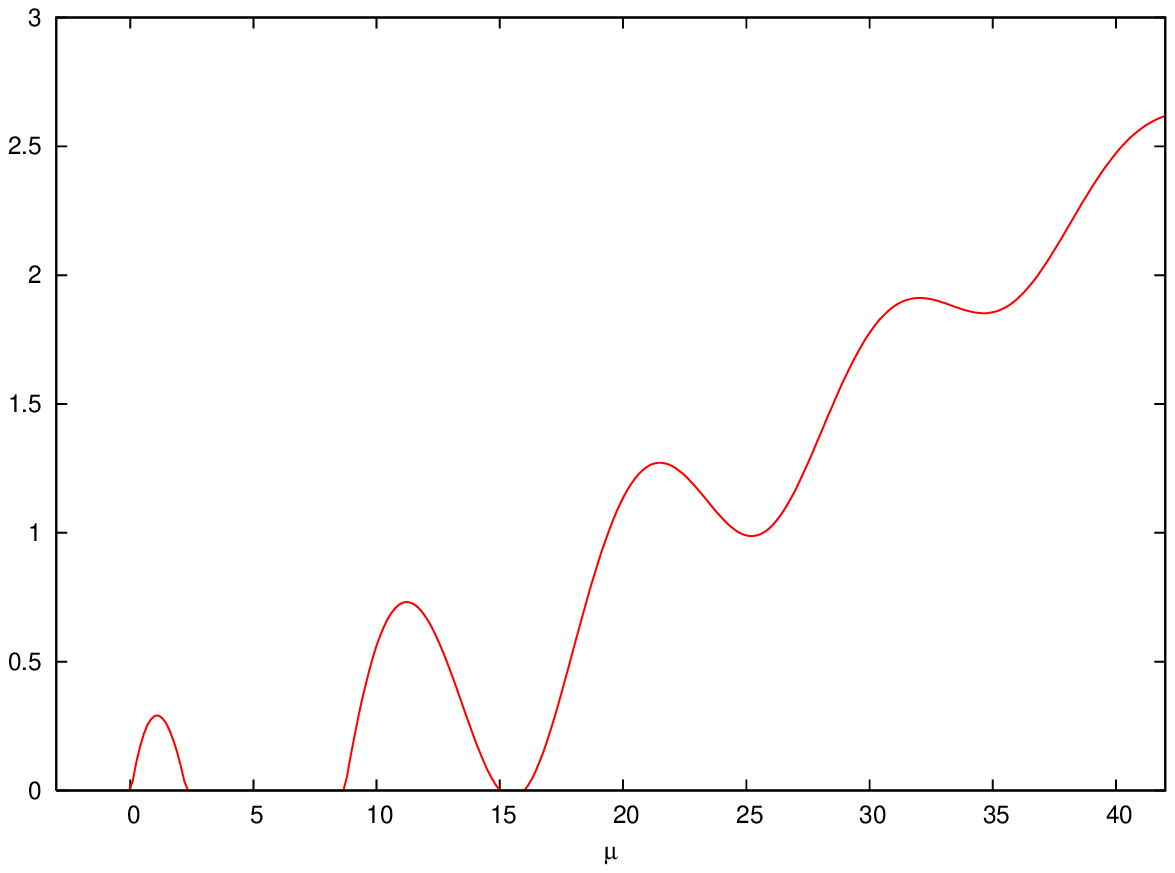}
  \end{center}
  \caption{\it Plot of the condensate density $r_\mu^2 ( =\rhocond)$ versus $\mu$, for $\beta=4$, $\lambda=5$.}
  \label{fig1}
\end{figure}

The properties of this model were then obtained numerically by finding this maximal solution of
the Euler-Lagrange equation and then evaluating the pressure using (\ref{pressure}).
As may be seen from Fig \ref{fig1}, for sufficiently large $\beta$, there may exist several critical 
values of $\mu$ which correspond to intervals of $r_\mu = 0$ and $r_\mu > 0$.

In addition Bru and Dorlas showed that Theorem \ref{thm_soln} holds in the presence of the gauge-symmetry 
breaking term. In that case, the corresponding Euler-Lagrange equation has a unique non-zero solution 
$r_\mu(\nu)$.

In this paper we shall analyse the cycle statistics of this model. $H_\sV$ is a grand-canonical 
Hamiltonian given by (\ref{I-R}) acting upon the bosonic Fock space $\FFsym$. Let $\Ham$ be the restriction of $H_\sV$
to the $n$ particle space $\Hcan{n}$, and its corresponding symmetrised subspace $\Hcansym{n}$. Then the 
grand-canonical partition function for this model may be written as
\begin{align*}
	\gPart 
&=	\trace_{\FFsym} \left[ \e^{-\beta (\gHam - \mu N_\sV)} \right]
=	\sum_{n=0}^\infty \trace_{\Hcansym{n}} \left[\e^{-\beta (\Ham - \mu n)} \right] 
\\
&=	\sum_{n=0}^\infty \frac{1}{n!} \sum_{\pi \in S_n} 
	\trace_{\Hcan{n}} \left[ U_\pi \e^{-\beta (\Ham - \mu n)} \right].
\end{align*}
where $S_n$ is the set of all permutations of $n$ items, and $U_\pi$ the unitary representation
of a permutation $\pi \in S_n$.

There is a natural probability measure (see \cite{DMP}) on the set of all
permutations $\bigcup_{n=0}^\infty S_n$ (taking $S_0 = \{ 1\}$) defined as
\[
	\gProb(\pi) 
=	\frac{1}{\gPart} \sum_{n=0}^\infty \frac{1}{n!} \trace_{\Hcan{n}} 
	\left[ U_\pi e^{-\beta (\Ham - \mu n)} \right] \mathcal{I}_{S_n}(\pi)
\]
where $\mathcal{I}$ is the indicator function, ensuring $\pi \in S_n$ for some $n$.

From the random walk formulation (see for example \cite{Toth}) one can see that the kernel of $\e^{-\beta \Ham}$
is positive and therefore the righthand side of this expression is positive.

Each permutation $\pi \in S_n$ can be decomposed uniquely into a number of cyclic permutations of lengths
$q_1, q_2, \dots, q_j$ with $j \le n$ and $q_1 + q_2 + \dots + q_j = n$.
For $q \in \{1,2,\ldots,n\}$, let $N_q(\pi)$ be the random variable
corresponding to the number of cycles of length $q$ in $\pi$. Then
the expectation of the number of $q$-cycles in the grand canonical ensemble is
\[
	\gEx(N_q) 
=	\sum_{j=0}^\infty j \gProb(N_q\!=\!j) 
\]
and the average density of particles in $q$-cycles is
\[
	c^\mu_\sV(q) = \frac{q \; \gEx(N_q) }{V}.
\]
This brings us then to the following definition.
\begin{definition}
The expected density of particles on cycles of \textbf{finite} length is given by
\begin{equation}
	\rhoshort = \lim_{Q \to \infty} \;\thermlim \; \sum_{q=1}^Q c_\sV^\mu(q)
\end{equation}
and the expected density of particles on cycles of \textbf{infinite} length is given by
\begin{equation}
	\rholong = \lim_{Q \to \infty} \;\thermlim \; \sum_{q=Q+1}^\infty c_\sV^\mu(q).
\end{equation}
Clearly $\rhodens = \rhoshort + \rholong$.
\end{definition}

For brevity, denote $c^\mu_{\phantom{l}}(q) = \thermlim c^\mu_\sV(q)$.
It is clearly easier to deal with $\rhoshort$ since we can take the thermodynamic limit 
inside the sum over $q$ to get
\[
	\rhoshort = \sum_{q=1}^\infty \; c^\mu_{\phantom{l}}(q).
\]
For the free Bose gas, the mean-field and the perturbed mean-field Bose gas, it has been
shown that $\rholong=\rhocond$, the condensate density. 
However in the case of the Hard-Core Infinite-Range-Hopping Bose-Hubbard model (see \cite{BolandPule}),
a different conclusion was obtained: that $\rholong > 0$ if and only if $\rhocond > 0$, 
but that in the presence of condensation these quantities were not necessarily equal.
We wish to argue that this is also the case for the chosen model.

We now state our results.
\begin{theorem}							\label{thm-short-cycles}
The density of cycles of finite length in the IRH Bose-Hubbard model may be expressed as
\[
	\rhoshort
=	\sum_{q=1}^\infty \e^{-\beta(q-\mu)q} \thermlim 
	\frac{ \trace_{\FFsym} \exp\big\{ -\beta ( 2\lambda q n_1 + H_\sV - \mu N_\sV ) \big\} }
	{ \trace_{\FFsym} \exp\big\{ -\beta ( H_\sV - \mu N_\sV ) \big\} }.
\]
where $n_1$ is an operator which counts the number of bosons on the site labelled $1$.
\end{theorem}

If one were to substitute $H_\sV$ for the approximating Hamiltonian $H^{\text{APP}}_\sV(r_\mu)$ (where again $r_\mu$ is 
the maximal solution of the Euler-Lagrange equation (\ref{euler-lagrange})) into the right hand side of this expression,
one would obtain:
\begin{equation}							\label{conj-short-cycles-approxed} 
	\rhoshort
=	\sum_{q=1}^\infty 
	\frac{  \trace_{\mathcal{F}_{+}(\CC)} \e^{ -\beta h_q(r_\mu) }}
	{ \trace_{\mathcal{F}_{+}(\CC)} \e^{ -\beta h_0(r_\mu)}}.
\end{equation}
where 
\[
	h_q(r) \vcentcolon= (1-\mu)(n + q) + \lambda (n + q)(n + q - 1) - r (a + a^\ast)
\]
is another single-site Hamiltonian (note that $h_0(r) = h(r)$).
This leads to the conjecture that (\ref{conj-short-cycles-approxed}) gives the
correct expression for $\rhoshort$.

Moreover the fact that a state corresponding to $H_\sV$ in the thermodynamic limit may be shown to be a convex
combination of one-site product states of the form
\[
	\omega(A) = \frac{\trace_{\mathcal{F}_{+}(\CC)} \e^{-\beta h_q(r_\mu) } A}
	{\trace_{\mathcal{F}_{+}(\CC)} \e^{-\beta h_0(r_\mu)}}
\]
supports this conjecture. We prove the conjecture for those values of $\mu$ such that $r_\mu=0$, 
but unfortunately are unable to do so when $r_\mu>0$. However we can prove 
a slightly weaker result with the addition of a gauge-symmetry breaking term.

Let $c_\sV^\mu(q,\nu)$ be the density of particles on cycles of length $q$ for 
the gauge-symmetry broken Hamiltonian $H_\sV(\nu)$.

\begin{theorem}								\label{prop3-equiv}
For $\mu$ such that $r_\mu=0$, we have 
\[
	c^\mu_{\phantom{l}}(q)
= 	\frac{  \trace_{\mathcal{F}_{+}(\CC)} \e^{ -\beta h_q(0) }}
	{ \trace_{\mathcal{F}_{+}(\CC)} \e^{ -\beta h_0(0) }}.
\]

More generally for any $\mu \in \RR$, for a fixed $\nu>0$ there exists a sequence $\nu_\sV \to \nu$ as $V \to \infty$,
independent of $q$ such that
\begin{equation}								\label{prop3-gaugebroken}
	\thermlim c_\sV^\mu(q,\nu_\sV)
=	\frac{ \trace_{\mathcal{F}_{+}(\CC)} \e^{ -\beta [ h_q(r_\mu(\nu)) - \nu (a + a^\ast) ] }}
	 { \trace_{\mathcal{F}_{+}(\CC)} \e^{ -\beta [ h_0(r_\mu(\nu)) - \nu (a + a^\ast) ] }}.
\end{equation}
\end{theorem}
Note that this theorem implies (\ref{conj-short-cycles-approxed}) for $r_\mu = 0$. 
This allows us to show that in the absence of condensation all particles are on short cycles. 

\begin{theorem}								\label{no-long}
In the absence of condensation, i.e. for those $\mu$ such that $r_\mu=0$, the density of particles 
on short cycles equals the total density of the system, that is:
\[
	\rhoshort = \rhodens \quad \Rightarrow \quad \rholong = 0.
\]
\end{theorem}
Moreover considering equation (\ref{prop3-gaugebroken}), if the thermodynamic limit and the limit 
removing the gauge-breaking source are interchangeable
then (\ref{conj-short-cycles-approxed}) follows for any $r_\mu$.

\begin{figure}[hbt]
\begin{center}
\includegraphics[width=16cm]{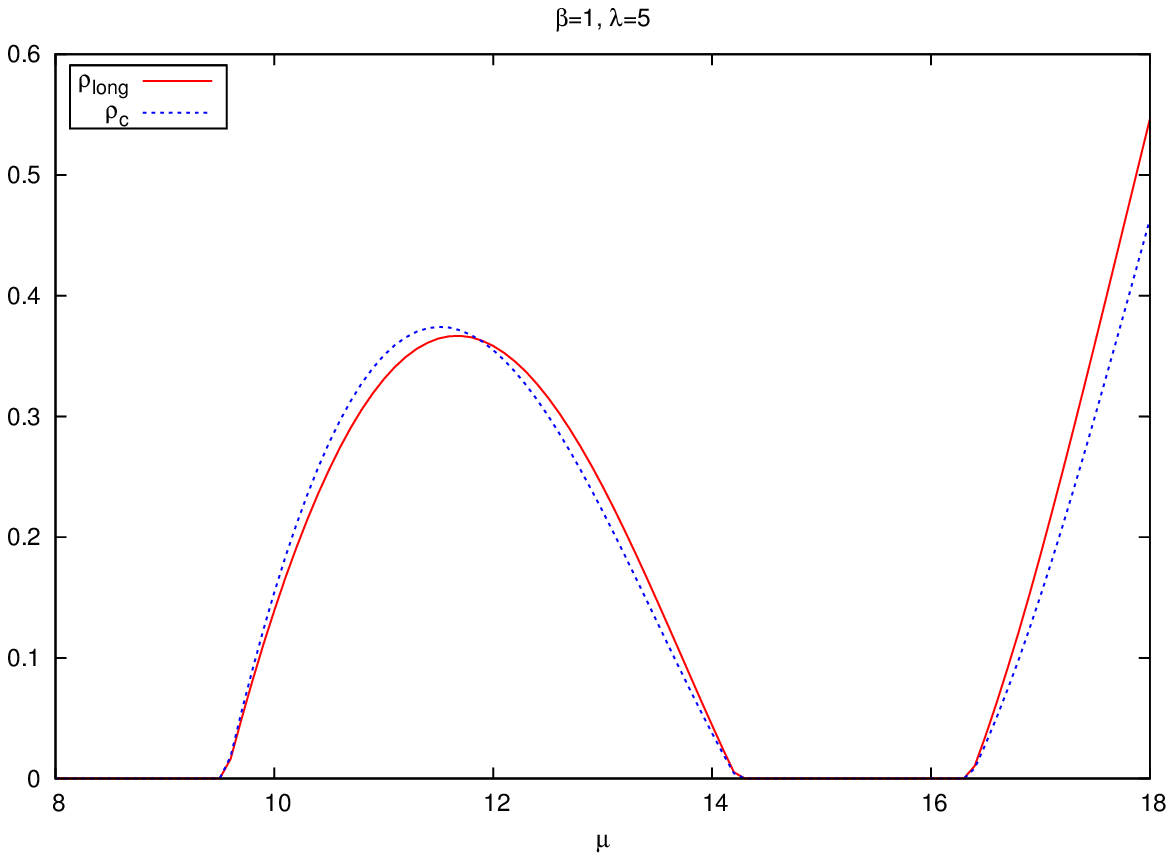}
\end{center}
\caption{\it Comparison of $\rholong$ with $\rhocond$}
\label{fig2}
\end{figure}

\begin{figure}
  \begin{center}
  \includegraphics[width=15.5cm]{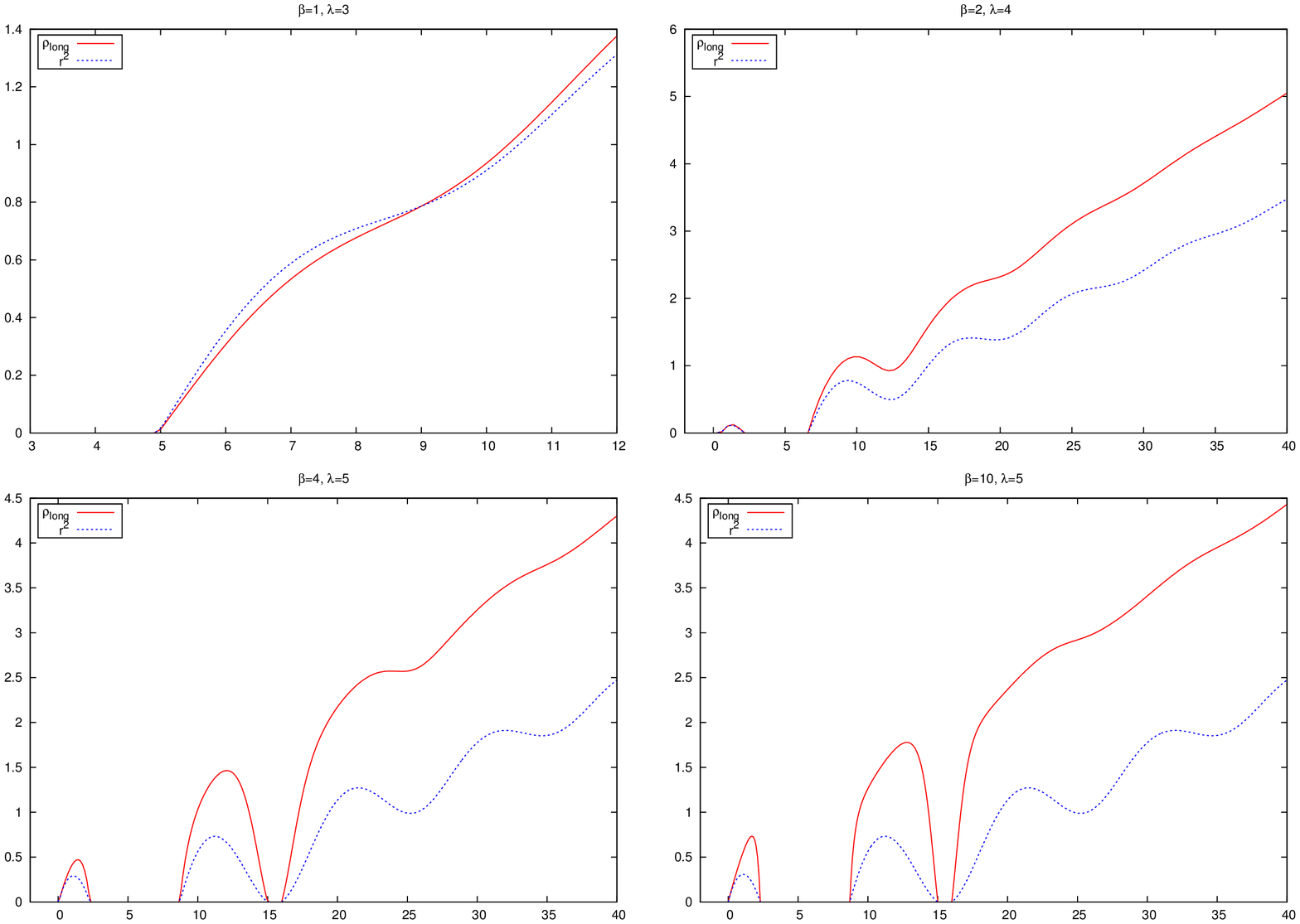}
    \caption{\it Comparison of $\rholong$ with $\rhocond$ for various values of $\beta$ and $\lambda$.}
    \label{test4}
  \end{center}
\end{figure}

Assuming the conjecture is correct, we apply some simple numerical techniques to (\ref{conj-short-cycles-approxed})
in order to compare long cycles with the Bose-Einstein condensate.
As may be seen from Figures \ref{fig2} and \ref{test4}
the calculations certainly agree with Theorem (\ref{no-long}), i.e. that the absence of condensation implies 
the lack of long cycles and visa versa. However more importantly they also indicate that while the 
presence of condensation coincides with the existence of long cycles, their respective densities are 
not necessarily equal. In fact, one can see that the long cycle density may be greater than 
or less than the condensate density for differing parameters.

\section{Proof of Theorem \ref{thm-short-cycles}}				\label{Section2}

Before proceeding to the study the cycle statistics for this model
we need to define the $n$-particle Hamiltonian in more detail. The Hilbert space for a single particle
on a lattice of $V$ sites is $\Hone\vcentcolon=\CC^\sV$ and on it we define the operator
\[
	h_\sV=I-P_\sV
\]
where $P_\sV$ is the orthogonal projection onto the unit vector
\[
	\mathbf{g}_\sV
=	\frac{1}{\sqrt{V}} \sum_{x=1}^V \mathbf{e}_x
=	\frac{1}{\sqrt{V}} (1,1,\dots,1)\in \Hone,
\]
with $\{\mathbf{e}_x\}_{x=1}^V$ the usual orthonormal basis for $\Hone$. 
$h_\sV$ is the orthogonal projection onto the subspace orthogonal to $\mathbf{g}_\sV$.
For an operator $A$ on $\Hone$, we define $A^{(n)}$ on
$\Hcan{n} \vcentcolon= \underbrace{\Hone \otimes \Hone \otimes \dots
\otimes \Hone}_{n \text{ times}}$, by \vspace{-0.2cm}
\[
    	A^{(n)}
=	A\otimes I\otimes \ldots \otimes I+I\otimes A\otimes \ldots \otimes I+\ldots
	+I\otimes I\otimes \ldots \otimes A.
\]
Let $\FF = \bigoplus_{n=0}^\infty \Hcan{n}$ denote the \textit{unsymmetrised} Fock space of $\Hone$
and define $d\Gamma(A)$ on $\FF$ as
\[
	d\Gamma(A) = \sum_{n=0}^\infty A^{(n)}
\]
where $\Hone^{(0)} \vcentcolon= \CC$ and $\Hone^{(1)} \vcentcolon= \Hone$.
With this notation we can write the free Hamiltonian acting on $\FF$ as:
\[
	H_\sV^{\text{free}} = d\Gamma(h_\sV).
\]
This represents a collection of particles on a lattice 
of $V$ sites which hop freely from site to site with no inter-particle or external interactions.
The hopping action is reflected by the $P_\sV$ operator.

For bosons we have to consider the symmetric subspace of $\FF$.
The symmetrisation projection $\sigma_{+}^{n}$ on $\Hcan{n}$ is defined by
\begin{equation}								\label{sym}
	\sigma_{+}^{n} = \frac{1}{n!} \sum_{\pi \in S_n} U_\pi
\end{equation}
where $U_\pi: \Hcan{n} \mapsto \Hcan{n}$ is the unitary representation of the permutation group
$S_n$ on $\Hcan{n}$ defined by
\[
	U_\pi(\phi_1 \otimes \phi_2 \otimes \cdots \otimes \phi_n)
=	\phi_{\pi(1)} \otimes \phi_{\pi(2)} \otimes \cdots
	\otimes \phi_{\pi(n)}, \;\; \phi_j \in \Hone, \; j=1, \dots, n; \;  \pi \in S_n.
\]
The symmetric $n$-particle subspace is $\Hcansym{n} \vcentcolon= \sigma_{+}^n \Hcan{n}$, allowing us to
define the \textit{symmetrised} Fock space as $\FFsym \vcentcolon= \bigoplus_{n=0}^\infty \Hcansym{n}$.

The operator which counts the number of particles at site $x$, $n_x$, is defined by
$n_x = d\Gamma(P_x) \vcentcolon= d\Gamma( |\mathbf{e}_x \ra\la \mathbf{e}_x |)$. Then
the total number operator is $N_\sV = \sum_{x=1}^\sV n_x$.

Let us define the Hamiltonian $\gHam$ on $\FF$ by
\begin{equation}										\label{B-H-notation}
	\gHam = d\Gamma(h_\sV) + \lambda \sum_{x=1}^\sV n_x (n_x - 1).
\end{equation}
This Hamiltonian restricted to $\FFsym$ is in fact the Bose-Hubbard Hamiltonian.


The proof of Theorem \ref{thm-short-cycles} is in three steps. First we obtain a 
convenient expression for $c_\sV^\mu(q)$, the density of particles on a cycle of length $q$ (Lemma \ref{Prop1}). 
This involves the partition function of $q$ distinguishable particles interacting with the boson system
through the Hamiltonian (\ref{B-H-notation}).
Then we construct a modified cycle density, denoted $\widetilde{c}\,_\sV^\mu(q)$, which neglects 
the hopping of the $q$ distinguishable particles and show that these cycle densities are equivalent in the limit
(Lemma \ref{Prop2}). Finally we simplify $\widetilde{c}\,_\sV^\mu(q)$ (Lemma \ref{Cor1}).

We shall denote the unitary representation of a $q$-cycle by $U_q: \Hcan{q} \to \Hcan{q}$, that is
\[
	U_q ( \phi_{i_1} \otimes \dots \otimes \phi_{i_q}) =
	\phi_{i_2} \otimes \dots \phi_{i_q} \otimes \phi_{i_1} .
\]
Denote the identity operator on $\Hcan{q}$ by $I^{(q)}$, and upon $\FFsym$ by $\mathbb{I}$ .
When there is no ambiguity we shall simply write $U_q$ for $U_q \otimes 
\mathbb{I}: \FFsym \to \FFsym$. Note that $[U_q, \sigma^n_{+}] = 0$. 

We state three lemmas without proof in the course of the argument of Theorem \ref{thm-short-cycles} 
and prove them shortly afterwards.

\begin{lemma}							\label{Prop1}
The density of particles on cycles of length $q$ is
\[
	c_\sV^\mu(q) 
=	\frac{1}{\gPart V} \trace_{\mathcal{H}_{q, \sV} }
	\left[ U_q \e^{-\beta ( \gHam - \mu N_\sV)} \right]
\]
where $\gPart$ is the grand-canonical partition function for $\gHam$ and
$\qFFsym \vcentcolon= \Hcan{q} \otimes \FFsym$.
\end{lemma}

This lemma uses cycle statistics to split the symmetric Fock space $\FFsym$ into the 
tensor product of two spaces, an unsymmetrised $q$-particle space $\Hcan{q}$
and a symmetrised Fock space $\FFsym$. Write
\begin{alignat*}{4}
	A^{(q)} &= A^{(q)} \otimes \mathbb{I}
	\qquad&\text{and}&\qquad
	&d\Gamma'(A) &= I^{(q)} \otimes d\Gamma(A)
\intertext{for any operator $A$ on $\Hone$. In this fashion, the number operators 
applied to $\mathcal{H}_{q, \sV}$ are defined as}
	N_x &= P_x^{(q)} \otimes \mathbb{I}
	\qquad&\text{and}&
	&n_x &= I^{(q)} \otimes d\Gamma(P_x).
\end{alignat*}
Then we may define $\gHam^{(q)}$ on $\mathcal{H}_{q, \sV}$ by
\[
	\gHam^{(q)} = h_\sV^{(q)} + d\Gamma'(h_\sV) + 
			\lambda \sum_{x=1}^\sV (n_x + N_x)(n_x + N_x - 1).
\]
Define a modified Hamiltonian which neglects the hopping of the $q$ distinguishable particles as follows:
\[
	\widetilde{H}_{\sV}^{(q)} = I^{(q)} +
		d\Gamma'(h_\sV) + \lambda \sum_{x=1}^\sV (n_x + N_x)(n_x + N_x - 1)
\]
so that $\gHam^{(q)} = \widetilde{H}_{\sV}^{(q)} - P_\sV^{(q)}$, and define the corresponding cycle density 
(henceforth called the ``modified cycle density'') by
\[
	\widetilde{c}\,_\sV^\mu(q) 
	= \frac{1}{\gPart} \frac{1}{V} \trace_{\mathcal{H}_{q, \sV}}
		\bigg[ U_q \e^{-\beta (\widetilde{H}_{\sV}^{(q)} - \mu N_\sV) } \bigg].
\]

Then we have the estimate:
\begin{lemma} 								\label{Prop2}
\begin{equation*}
   	\left|c_\sV^\mu(q)-\widetilde{c}\,_\sV^\mu(q)\right |
\le 	\frac{\e^{\beta q\mu}( 1 - \e^{-\beta q})}{V}.
\end{equation*}
\end{lemma}
This implies that in the thermodynamic limit, we are able to disregard the hopping of the 
$q$-unsymmetrised particles in the cycle density.

The modified cycle density can be re-expressed as:
\begin{lemma}								\label{Cor1}
\[
	\widetilde{c}\,_\sV^\mu(q) 
= 	\e^{-\beta(q-\mu)q}  
	\frac{ \trace_{\FFsym} \exp\big\{-\beta ( 2\lambda q n_1 + H_\sV - \mu N_\sV ) \big\} }
	{ \trace_{\FFsym} \exp\big\{-\beta ( H_\sV -\mu N_\sV )\big\} }.
\]
where $n_1$ is an operator which counts the number of bosons on the site labelled 1 of the lattice.
\end{lemma}

Combining the above information, we deduce that
\[
	c^\mu_{\phantom{l}}(q)
=	\e^{-\beta(q-\mu)q} \thermlim 
	\frac{ \trace_{\FFsym} \exp\big\{-\beta (  2\lambda q n_1 + H_\sV -\mu N_\sV ) \big\} }
	{ \trace_{\FFsym} \exp\big\{-\beta ( H_\sV -\mu N_\sV ) \big\} }
\]
and this completes the proof of the Theorem.

Now we shall prove the lemmas.

\subsection{Proof of Lemma \ref{Prop1}}
The canonical expectation of the number of $q$-cycles may be found to be
\begin{equation*}
	\cEx(N_q) 
=	\frac{1}{\cPart} \frac{1}{q}
	\trace_{\Hcan{q} \otimes \Hcansym{n-q}}
	\left[ (U_q \otimes I^{(n-q)}) \e^{-\beta \Ham} \right]
\end{equation*}
by following the proof of Proposition $3.1$ of the preceding paper\cite{BolandPule} and omitting all the hard-core
projections.

Then going to the grand-canonical ensemble we obtain:
\begin{align*}
\gEx(N_q) 
    &= \sum_{n=q}^\infty \frac{e^{\beta \mu n} \cPart \cEx(N_q)}{\gPart} 
\\
    &=  \sum_{n=q}^\infty \frac{1}{q \, \gPart} 
		\trace_{ \Hcan{q} \otimes \Hcansym{n-q}} 
		\left[ (U_q \otimes I^{(n-q)}) \e^{-\beta(\Ham-\mu n)} \right]
\\
    &=  \frac{1}{q \, \gPart} \trace_{ \Hcan{q} \otimes \FFsym }
		\left[ (U_q \otimes \mathbb{I}) \e^{-\beta(\gHam -\mu N_\sV)} \right].
\end{align*}
Hence
\[
 	c_\sV^\mu(q) 
= 	\frac{q \; \gEx(N_q) }{V}
=	\frac{1}{\gPart V} \trace_{ \Hcan{q} \otimes \FFsym }
		\left[ (U_q \otimes \mathbb{I}) \e^{-\beta(\gHam -\mu N_\sV)} \right]
\]
as desired.
\hfill $\square$

\subsection{Proof of Lemma \ref{Prop2}}

The following technique has been employed to prove a similar result in \cite{BolandPule}.
However in this case there are several important differences and therefore we give the proof in full.

To prove Lemma \ref{Prop2} we have to obtain an upper bound for
\[
	\left |\trace_{\qFFsym}
	\Big[ U_q \e^{-\beta (\gHam - \mu N_\sV) } \Big]
	- \trace_{\qFFsym}
	\Big[ U_q \e^{-\beta (\widetilde{H}_\sV^{(q)} - \mu N_\sV) } \Big]\right |.
\]
In order to do this we first shall introduce some notation.
Let $\{ \phi_k \}_{k=0}^\infty$ be an orthonormal basis for $\FFsym$.

Let $\Lambda_{\sV}^{(q)}$ be the set of ordered $q$-tuples of (not necessarily distinct) indices 
of $\Lambda_\sV$ and for $\mathbf{i}=(i_1, i_2, \dots , i_q) \in \Lambda_{\sV}^{(q)}$
let
\[
	|\ii\ra = | i_1, i_2, \dots , i_q \ra = \mathbf{e}_{i_1} \otimes \mathbf{e}_{i_2}
	\otimes \dots \otimes \mathbf{e}_{i_q}.
\]
Then $\{|\ii \ra\,|\,\mathbf{i}\in \Lambda_{\sV}^{(q)}\}$ is an orthonormal basis for $\Hcan{q}$.

A basis for $\qFFsym$ may therefore be formed by taking the tensor product of the bases
of $\Hcan{q}$ and $\FFsym$, so the set
$\{|\mathbf{i} \ra \otimes \phi_k  \,|\, k=1,2,\dots ; \mathbf{i}\in \Lambda_{\sV}^{(q)} \}$
is an orthonormal basis for $\qFFsym$. For brevity we shall write
\begin{equation}									\label{basis-hvq}
	| \ii; k \ra = |\mathbf{i} \ra \otimes \phi_k.
\end{equation}

For simplicity, denote $P \vcentcolon= P_\sV^{(q)}$ and $H \vcentcolon= \widetilde{H}_{\sV}^{(q)} - \mu N_\sV$.
We expand
\[
	\trace_{\qFFsym} \left[U_q \e^{-\beta (\gHam - \mu N_\sV )} \right]
=	\trace_{\qFFsym} \left[U_q \e^{-\beta (H - P) } \right]
\]
in a Dyson series in powers of $P$. If $m\ge1$, the $m^\text{th}$ term of this series is
\begin{multline}									\label{mth-term-dyson}
	X_m
=	\beta^m \int_0^1\hskip -0.3cm ds_1 \int_0^{s_1} \hskip -0.4cm ds_2
	\dots  \int_0^{s_{m-1}}\hskip -0.8cm ds_m \;
	\trace_{\qFFsym} \bigg[ \e^{-\beta H (1-s_1)} P \e^{-\beta H (s_1-s_2)} P\cdots
\\
	\cdots P \e^{-\beta H (s_{m-1}-s_m)} P \e^{-\beta H s_m} U_q \bigg]. \qquad
\end{multline}
Let $P_r = I \otimes \dots \otimes \underbrace{P_\sV}_{r^{\text{th}} \text{place}} \otimes \dots \otimes I$,
so that $P = \sum_{r=1}^q P_r$. Then  \vspace{-0.4cm}
\[
	X_m = \beta^m \sum_{r_1=1}^q \sum_{r_2=1}^q \dots \sum_{r_m=1}^q X_m( r_1, r_2, \dots , r_m )
\]
where 
\begin{multline}									\label{mth-term-dyson-r}
	X_m( r_1, r_2, \dots , r_m )
=	\int_0^1\hskip -0.3cm ds_1 \int_0^{s_1} \hskip -0.4cm ds_2
	\dots  \int_0^{s_{m-1}}\hskip -0.8cm ds_m \;
	\trace_{\qFFsym} \bigg[ \e^{-\beta H (1-s_1)} P_{r_1} \e^{-\beta H (s_1-s_2)} P_{r_2} \cdots
\\
	\cdots P_{r_{m-1}} \e^{-\beta H (s_{m-1}-s_m)} P_{r_m} \e^{-\beta H s_m} U_q \bigg]. \qquad
\end{multline}
In terms of (\ref{basis-hvq}), the basis of $\qFFsym$, we may write
\begin{multline}									\label{trace_term}
	X_m( r_1, r_2, \dots , r_m )
=	\int_0^1\hskip -0.3cm ds_1 \int_0^{s_1} \hskip -0.4cm ds_2
	\dots  \int_0^{s_{m-1}}\hskip -0.8cm ds_m \;\;
	\sum_{k^0, \dots\ ,k^m} \;\; \sum_{\ii^0} \cdots
	\sum_{\ii^m }
\\
	\qquad\qquad
	\la \ii^0; k^0 | \e^{-\beta H (1-s_1)} P_{r_1}
	| \ii^1; k^1 \ra
	\la \ii^1; k^1 | \e^{-\beta H (s_1-s_2)} P_{r_2}
	| \ii^2; k^2 \ra \cdots
\\
	\cdots
	\la \ii^{m-1}; k^{m-1} |\e^{-\beta H (s_{m-1} -s_m)} P_{r_m}
	| \ii^m; k^m \ra
	\la \ii^m; k^m |\e^{-\beta H s_m}  U_q  | \ii^0; k^0 \ra
\end{multline}
where it is understood that the $\ii$ summations are over $\Lambda_{\sV}^{(q)}$,
the set of ordered $q$-tuples (not necessarily distinct) of $\Lambda_\sV$,
and the $k$ summations are over the bases for $\FFsym$.

Notice that we may express
\[
	\e^{-\beta H s} | \ii; k \ra
= 	\e^{-\beta q (1-\mu) s} \; | \ii; \e^{-\beta H^\ii s } | k \ra
\]
where
\[
	H^\ii = \gHam - \mu N_\sV + \lambda \sum_{x=1}^V ( N^\ii_x ( N^\ii_x - 1) + 2N^\ii_x n_x )
\]
and $N^\ii_x = \sum_{j=1}^q \delta_{x, i_j}$ counts the number of particles at site $x$ which are in $q$-space.
Also, for any fixed $r$:
\begin{equation}							\label{r-hopper}
	P_r | \ii; k \ra 
= 	\frac{1}{V} \sum_{j=1}^V | i_1,\dots,\widehat{i_r}, j,\dots,i_q ; k \ra
\end{equation}
where the hat symbol implies that the term is removed from the sequence.

It is convenient to define the operation $[r,x](\ii)$ which inserts the value of $x$
in the $r^\text{th}$ position of $\ii$ instead of $i_r$.
So for example taking the ordered triplet $\ii = (5,4,1)$,
then $[2,8](\ii) = (5,8,1)$. We shall denote the composition of these
operators as $[r_k, x_k; \, \dots \, ; r_2, x_2 ; r_1, x_1] \vcentcolon= [r_k, x_k] \circ \cdots \circ [r_2, x_2] \circ [r_1, x_1]$.

So (\ref{r-hopper}) may be written as
\[
	P_r | \ii; k \ra 
= 	\frac{1}{V} \sum_{j=1}^V | [r,j](\ii) ; k \ra.
\]

Using these facts, a single inner product term of (\ref{trace_term}) may be expressed as
\begin{align*}
	\la \ii ; k | \e^{-\beta H s} P_r | \jj; k' \ra
&=	\e^{-\beta q (1-\mu) s} \; \la k |
	\e^{-\beta H^\ii s } | k' \ra \; \la \ii | P_r| \jj \ra
\\
&=	\frac{\e^{-\beta q (1-\mu) s}}{V} \la k | \e^{-\beta H^\ii s } | k' \ra
	\sum_{m=1}^V \la \ii | j_1, \dots \widehat{j_r}, m, \dots, j_q \ra 
\\
&=	\frac{\e^{-\beta q (1-\mu) s}}{V} \la k | \e^{-\beta H^\ii s } | k' \ra
 	\sum_{m=1}^V \delta_{i_1, j_1} \dots \widehat{\delta_{i_r, j_r}} \delta_{i_r, m} \dots \delta_{i_q, j_q}
\\
&=	\frac{\e^{-\beta q (1-\mu) s}}{V} \la k | \e^{-\beta H^\ii s } | k' \ra \;
	\delta_{i_1, j_1} \dots \widehat{\delta_{i_r, j_r}} \dots \delta_{i_q, j_q}.
\end{align*}

Now if we sum over $\jj$
\begin{align*}
	\sum_{\jj} \la \ii ; k | \e^{-\beta H s} P_r | \jj; k' \ra \la \jj; k' |
&=	\frac{\e^{-\beta q (1-\mu) s}}{V} \la k | \e^{-\beta H^\ii s } | k' \ra 
	\sum_{j_r=1}^V \la i_1, \dots, \widehat{i_{r}}, j_r, \dots, i_q; k' |
\\
&=	\frac{\e^{-\beta q(1-\mu) s}}{V} \la k | \e^{-\beta H^\ii s } | k' \ra 
	\sum_{j_r=1}^V \la [r, j_r](\ii) ; k' |.
\end{align*}

Performing two summations for fixed $r_1$ and $r_2$ we get:
\begin{align*}
&	\sum_{\ii^1} \sum_{\ii^2} \la \ii^0 ; k^0 | \e^{-\beta H s} P_{r_1} | \ii^1; k^1 \ra \;
	\la \ii^1; k^1 | \e^{-\beta H t} P_{r_2} | \ii^2; k^2 \ra \; \la \ii^2; k^2 |
\\
&\quad= 	\frac{\e^{-\beta q (1-\mu) s}}{V} \sum_{i^1_{r_1}=1}^V \sum_{\ii^2} 
	\la k^0 | \e^{-\beta H^{\ii^0} s } | k^1 \ra \; \la [r_1, i^1_{r_1}](\ii^0) ; k^1 |
	 \e^{-\beta H t} P_{r_2} | \ii^2; k^2 \ra \; \la \ii^2; k^2 |
\\
&\quad=	\frac{\e^{-\beta q (1-\mu) (s+t)}}{V^2} \sum_{i^1_{r_1}=1}^V \sum_{i^2_{r_2}=1}^V 
	\la k^0 | \e^{-\beta H^{\ii^0} s } | k^1 \ra \; \la k^1 | \e^{-\beta H^{[r_1, i_{r_1}](\ii^0)} t } | k^2 \ra \;
	\la [r_2, i^2_{r_2} ; r_1, i^1_{r_1} ](\ii^0) ; k^2 |.
\end{align*}

Thus (\ref{mth-term-dyson-r}) looks like
\begin{align*}
  	X_m&( r_1, r_2, \dots , r_m )
\\
=& 	\frac{\e^{-\beta q(1-\mu)}}{V^m} 
	\int_0^1\hskip -0.3cm ds_1 \int_0^{s_1} \hskip -0.4cm ds_2
	\dots  \int_0^{s_{m-1}}\hskip -0.8cm ds_m \;\;
	\sum_{k^0 \dots k^m} \sum_{\ii^0}
	\sum_{i^1_{r_1} = 1}^V
	\sum_{i^2_{r_2} = 1}^V
	\cdots \sum_{i^m_{r_m} =1}^V
	\la k^0 |  \e^{-\beta(1-s_1) H^{\ii^0} }
	| k^1 \ra
\\
&	\qquad \la k^1 |
	\e^{-\beta (s_1-s_2) H^{[r_1,i^1_{r_1}](\ii^0)}}
	| k^2 \ra
	\la k^2 | \e^{-\beta (s_2-s_3)
	H^{[r_2, i^2_{r_2} ; r_1, i^1_{r_1}](\ii^0)} }
	| k^3 \ra
	\cdots
\\
&	\qquad \cdots
	\la k^m |
	\e^{-\beta s_m H^{ [r_m, i^m_{r_m}; \, \dots \, ; r_2, i^2_{r_2} ; r_1, i^1_{r_1}](\ii^0)} }
	| k^0 \ra
	\la [r_m, i^m_{r_m}; \, \dots \, ; r_2, i^2_{r_2} ; r_1, i^1_{r_1}](\ii^0) | U_q \ii^0 \ra
\\[0.4cm]
=& 	\frac{\e^{-\beta q(1-\mu)}}{V^m}
	\int_0^1\hskip -0.3cm ds_1 \int_0^{s_1} \hskip -0.4cm ds_2
	\dots  \int_0^{s_{m-1}}\hskip -0.8cm ds_m
\\
&	\qquad \sum_{\ii^0}
	\sum_{i^1_{r_1} = 1}^V
	\sum_{i^2_{r_2} = 1}^V
	\cdots \sum_{i^m_{r_m} = 1}^V
	\la [r_m, i^m_{r_m}; \, \dots \, ; r_2, i^2_{r_2} ; r_1, i^1_{r_1}](\ii^0) | U_q \ii^0 \ra
\\
&	\qquad \trace_{\FFsym} \Bigg[ \e^{-\beta (1-s_1) H^{\ii^0} }
	\e^{-\beta (s_1-s_2) H^{[r_1, i^1_{r_1}](\ii^0)} }
	\cdots \cdots
	\e^{-\beta s_m H^{[r_m, i^m_{r_m}; \, \dots \, ; r_2, i^2_{r_2} ; r_1, i^1_{r_1}](\ii^0)} }
	\Bigg].
\end{align*}
From the H\"older inequality (see for example Manjegani \cite{Manjegani}), for 
non-negative trace class operators $A_1, A_2, \dots , A_{m+1}$ we have the inequality
\[
	\left| \trace \big( A_1 A_2 \dots A_{m+1} \big) \right|
\le	\trace \big| A_1 A_2 \dots A_{m+1} \big|
\le	\prod_{k=1}^{m+1} \big( \trace A_k^{p_k} \big)^{\tfrac{1}{p_k}}
\]
where $\sum_{k=1}^{m+1} \tfrac{1}{p_k} = 1$, $p_i \ge 1$.

Set $p_1 = \frac{1}{1-s_1},\ p_2 = \frac{1}{s_1 - s_2},\ \dots ,\ p_m = \frac{1}{s_{m-1}-s_m},
p_{m+1} = \frac{1}{s_m}$. Taking the modulus of the above trace
\begin{align*}
&	\Bigg| \trace_{\FFsym}  \Bigg[
	\e^{-\beta H^{\ii^0} (1-s_1)}
	\e^{-\beta H^{[r_1, i^1_{r_1}](\ii^0)} (s_1-s_2)}
	\; \cdots
	\cdots
	\e^{-\beta H^{[r_m, i^m_{r_m}; \, \dots \, ; r_2, i^2_{r_2} ; r_1, i^1_{r_1}](\ii^0)} (s_m)}
	\Bigg]  \Bigg|
\\
&\le	\quad \trace_{\FFsym} \bigg[
	\e^{-\beta H^{\ii^0}} \bigg]^{1-s_1}
	\trace_{\FFsym} \bigg[
	\e^{-\beta H^{[r_1, i^1_{r_1}](\ii^0)}} \bigg]^{s_1-s_2} \cdots
\\
&	\qquad\qquad\qquad \cdots \trace_{\FFsym}
	\bigg[
	\e^{-\beta H^{[r_m, i^m_{r_m}; \, \dots \, ; r_2, i^2_{r_2} ; r_1, i^1_{r_1}](\ii^0)}}
	\bigg]^{s_m} \!\!\!.
\end{align*}
Since the trace is independent of the $V-q$ sites $\{\ii^0, [r_1,i^1_{r_1}](\ii^0),\,  \dots,
\, [r_m, i^m_{r_m}; \, \dots\\ \, ; r_2, i^2_{r_2} ; r_1, i^1_{r_1}](\ii^0)\}$,
the product of all the trace terms above is equal to
\[
	\trace_{\FFsym} \left[ 
		\e^{-\beta H^{\boldl}} \right]
\]
with $\boldl = \{V-q+1, V-q+2, \dots ,V\}$.

This is independent of the $\ii^0$ and $i$ summations, so we need only consider
\begin{equation}  								\label{sum_of_cycled_inner_products} 
	\sum_{i^1_{r_1} = 1}^V
	\sum_{i^2_{r_2} = 1}^V
	\cdots \sum_{i^m_{r_m} = 1}^V
	\sum_{\ii^0}
	\la [r_m, i^m_{r_m}; \, \dots \, ; r_2, i^2_{r_2} ; r_1, i^1_{r_1}](\ii^0) | U_q \ii^0 \ra.
\end{equation}

Fix the values of $r_1, \dots, r_m$ and $i^1_{r_1}, i^2_{r_2}, \dots , i^m_{r_m}$. We intend to show that
\[
	\sum_{\ii^0}
	\la [r_m, i^m_{r_m}; \, \dots \, ; r_2, i^2_{r_2} ; r_1, i^1_{r_1}](\ii^0) | U_q \ii^0 \ra = 1.
\]
If $\{r_1, r_2, \dots , r_m\} \ne \{1,2,\dots, q\} $, 
then $| [r_m, i^m_{r_m}; \, \dots \, ; r_2, i^2_{r_2} ; r_1, i^1_{r_1}](\ii^0) \ra$ is of the form
\[
	| j_1, j_2, \dots ,j_{n_1},i^0_{n_1+1}, \dots ,i^0_{n_2}, j_{n_2+1},
	\dots, j_{n_3}, i^0_{n_3+1}, \dots  ,
	i^0_{n_4}, j_{n_4+1}, \dots \dots\ra
\]
where $\{n_1, n_2, \dots \}$ is a non-empty ordered set of distinct integers between 0 and $q$.
This vector is clearly orthogonal to $U_q \ii^0$ except for the single choice of
\[
	\ii^0 = |  j_2, \dots , j_{n_1-1}, j_{n_1}, \dots , j_{n_1}, j_{n_2+1},
	\dots, j_{n_3-1}, j_{n_3}, \dots ,
	j_{n_3}, j_{n_4+1}, \dots \dots , j_1\ra.
\]

For the case $\{r_1, r_2, \dots , r_m\} = \{1,2,\dots, q\}$ 
notice that $| [r_m, i^m_{r_m}; \, \dots \, ; r_2, i^2_{r_2} ; r_1, i^1_{r_1}](\ii^0) \ra$ is independent of
$\ii^0$ so we may take it to be
\[
	|[r_m, i^m_{r_m}; \, \dots \, ; r_2, i^2_{r_2} ; r_1, i^1_{r_1}](\mathbf{s}^0) \ra
\]
where $\mathbf{s}^0 = (1,2,3,\dots,q)$. For each choice of $i^1_{r_1}, i^2_{r_2}, \dots , i^m_{r_m}$ 
there exists only one possible $\ii^0 \in \Lambda_\sV^{(q)}$ such that
\[
	\la [r_m, i^m_{r_m}; \, \dots \, ; r_2, i^2_{r_2} ; r_1, i^1_{r_1}](\mathbf{s}^0) | U_q \ii^0 \ra \ne 0.
\]
So we may conclude that
\begin{equation}
	\sum_{\ii^0}
	\sum_{i^1_{r_1} = 1}^V
	\sum_{i^2_{r_2} = 1}^V
	\cdots \sum_{i^m_{r_m} = 1}^V
	\la [r_m, i^m_{r_m}; \, \dots \, ; r_2, i^2_{r_2} ; r_1, i^1_{r_1}](\ii^0) | U_q \ii^0 \ra
	= V^m
\end{equation}
and by using this, we see that the modulus of (\ref{mth-term-dyson-r}) may bounded above by
\begin{align*}
    |X_m( r_q, r_2, \dots, r_m)| & \le \trace_{\FFsym} \left[ 
		\e^{-\beta \Htilde^{\boldl}} \right]
		\e^{-\beta q(1-\mu)} \frac{1}{m! V^m}
\\
&   \qquad \times
	\sum_{\ii^0}
	\sum_{i^1_{r_1} = 1}^V
	\sum_{i^2_{r_2} = 1}^V
	\cdots \sum_{i^m_{r_m} = 1}^V
	\la [r_m, i^m_{r_m}; \, \dots \, ; r_2, i^2_{r_2} ; r_1, i^1_{r_1}](\ii^0) | U_q \ii^0 \ra
\\
&=	\trace_{\FFsym} \left[ 
		\e^{-\beta \Htilde^{\boldl}} \right] \frac{\e^{-\beta q(1-\mu)} }{m!}.
\end{align*}
which is independent of $r_1, r_2, \dots, r_m$.
Hence the modulus of (\ref{mth-term-dyson}), the $m^\text{th}$ term of the Dyson series, may be bounded above by
\begin{align*}
	| X_m | 
\enspace\le\enspace
	\beta^m \sum_{r_1=1}^q \cdots \sum_{r_m=1}^q |X_m( r_q, r_2, \dots, r_m)|
\enspace\le\enspace
	\trace_{\FFsym} \left[ 
		\e^{-\beta \Htilde^{\boldl}} \right]\e^{-\beta q(1-\mu)} \frac{q^m \beta^m}{m!}.
\end{align*}
Noting that the zeroth term of the Dyson series is
\[
	X_0 = \trace_{\mathcal{H}_{q,V}} \left[U_q \e^{-\beta H} \right] =
	\trace_{\mathcal{H}_{q,V}} \left[U_q \e^{-\beta ( \widetilde{H}_\sV^{(q)} - \mu N_\sV ) } \right],
\]
we may re-sum the series to obtain
\begin{align*}
&\Bigg| \trace_{\mathcal{H}_{q,\sV}}
            \left[U_q \e^{-\beta (H_{\sV} - \mu N_\sV) } \right]
        - \trace_{\mathcal{H}_{q,\sV}}
            \left[U_q \e^{-\beta ( \widetilde{H}_\sV^{(q)} - \mu N_\sV ) } \right]  \Bigg| \\
& \quad \le \trace_{\FFsym} \left[
	\e^{-\beta \Htilde^{\boldl}}  \right]\e^{-\beta q(1-\mu)} 
	\sum_{m=1}^\infty \frac{q^m \beta^m}{m!} \\
& \quad = \trace_{\FFsym} \left[
	\e^{-\beta \Htilde^{\boldl}}  \right]
	\e^{\beta q\mu} ( 1 - \e^{-\beta q}).
\end{align*}
Thus
\begin{align*}
   	\left|c_\sV^\mu(q)-\widetilde{c}\,_\sV^\mu(q)\right |
 &= 	\frac{1}{V} \left| \frac{
	\trace_{\mathcal{H}_{q,\sV}}
		\big[U_q \e^{-\beta ( H_{\sV} - \mu N_\sV) } \big]
	- \trace_{\mathcal{H}_{q,\sV}}
		\big[U_q \e^{-\beta ( \widetilde{H}_\sV^{(q)} - \mu N_\sV ) } \big]}{\gPart} \right|
\\
  &\le 	\frac{\e^{\beta q\mu} ( 1 - \e^{-\beta q})}{V} 
  	\frac{\trace_{\FFsym} \big[
		\e^{-\beta \Htilde^{\boldl}}  \big]}{\gPart}.
\end{align*}
Since $H_{\sV} - \mu N_\sV - \Htilde^{\boldl} = \lambda 
\sum_{x=1}^\sV ( N^\boldl_x ( N^\boldl_x - 1) + 2N^\boldl_x n_x ) \ge 0$,
the second fraction is not greater than 1, implying
\[
	\left|c_\sV^\mu(q)-\widetilde{c}\,_\sV^\mu(q)\right |
\le	\frac{\e^{\beta q\mu} ( 1 - \e^{-\beta q})}{V}
\]
which goes to zero in the limit $V \to \infty$, as desired.\hfill $\square$

\subsection{Proof of Lemma \ref{Cor1}}
The modified cycle density $\widetilde{c}\,_\sV^\mu(q)$ may be simplified as follows:
\begin{align*}
	\widetilde{c}\,_\sV^\mu(q)
=&	\frac{\e^{-\beta (1-\mu) q}}{V \gPart}
   	\sum_{i_1=1}^V \dots \sum_{i_q=1}^V \sum_{k=1}^\infty
	\la i_1, i_2, \dots, i_q; k |
\\
&	\quad\exp\left\{ -\beta \left( d\Gamma'(h_\sV)
		+ \lambda \sum_{x=1}^V (n_x + N_x)(n_x + N_x -1) - \mu \sum_{x=1}^V n_x
	\right) \right\}
\\
&	\quad U_q | i_1, i_2, \dots, i_q; k \ra
\\[0.3cm]
=&	\frac{\e^{-\beta (1-\mu) q}}{V \gPart}
   	\sum_{i_1=1}^V \dots \sum_{i_q=1}^V \sum_{k=1}^\infty
	\la i_1, i_2, \dots, i_q |  U_q | i_1, i_2, \dots, i_q \ra
\\
&	\quad \la k | \exp\left\{ -\beta \left( d\Gamma(h_\sV)
		+ \lambda \sum_{x=1}^V (n_x + \sum_{j=1}^q \delta_{i_j,x})
		(n_x + \sum_{j=1}^q \delta_{i_j,x} -1) - \mu \sum_{x=1}^V n_x
	\right) \right\} | k \ra
\intertext{and as $\la i_1, i_2, \dots, i_q | i_2, i_3, \dots, i_q, i_1 \ra \ne 0$ if and only if
$i_1 = i_2 = \dots = i_q \vcentcolon= i$ then}
=&	\frac{\e^{-\beta (1-\mu) q}}{V \gPart}
   	\sum_{i=1}^V \sum_{k=1}^\infty
	\la k |
\\
&	\quad	\exp\left\{ -\beta \left( d\Gamma(h_\sV)
		+ \lambda \sum_{x=1}^V (n_x + q \delta_{i x})
		(n_x + q \delta_{i x}  -1) - \mu \sum_{x=1}^V n_x \right) \right\}
	|k \ra
\\
=&	\frac{\e^{-\beta (1-\mu) q}}{V \gPart}
	\sum_{i=1}^V \trace_{\FFsym}
\\
&	\quad \exp\Bigg\{ -\beta \Bigg( d\Gamma(h_\sV)
		+ \lambda (n_i + q)(n_i + q  -1)
		+ \lambda \sum_{\stackrel{x=1}{x \ne i}}^V n_x ( n_x - 1)
		- \mu \sum_{x=1}^V n_x \Bigg) \Bigg\}
\\
=&	\frac{\e^{-\beta (1-\mu) q}}{\gPart}
	\trace_{\FFsym} \Bigg[
	\exp\Bigg\{ -\beta \Bigg( d\Gamma(h_\sV)
		+ \lambda (n_1 + q)(n_1 + q - 1)
\\
&	\quad
		+ \lambda \sum_{x=2}^V n_x ( n_x - 1)
		 - \mu \sum_{x=1}^V n_x \Bigg) \Bigg\}
	\Bigg]
\\
\end{align*}
since the trace is independent of the basis chosen. Hence we obtain
\[
	\widetilde{c}\,_\sV^\mu(q) 
= 	\e^{-\beta (q-\mu) q}
	\frac{ \trace_{\FFsym} \exp\big\{-\beta (2\lambda q n_1 + H_\sV -\mu N_\sV) \big\} }
	{ \trace_{\FFsym} \exp\big\{-\beta (H_\sV -\mu N_\sV) \big\} }.
\]\hfill $\square$

\section{Proof of Theorem \ref{prop3-equiv}}
For convenience, denote $\Hv{q} \vcentcolon= 2\lambda q n_1 + (q-\mu)q + H_\sV -\mu N_\sV$. 
Due to Lemmas \ref{Prop2} and \ref{Cor1}, we need to consider
\[
	c^\mu_{\phantom{l}}(q)
=	\thermlim \frac{\trace_{\FFsym} \big[\e^{- \beta \Hv{q}}\big]}
	{\trace_{\FFsym} \big[\e^{-\beta \Hv{0}}\big]}.
\]
Recall that
\begin{equation*}
	H_\sV
=	\frac{1}{2V}\!\!\!\sum_{x,y=1}^V (a^\ast_x-a^\ast_y)(a\astr_x-a\astr_y)
	+ \lambda \sum_{x=1}^V n_x (n_x-1).
\end{equation*}

Motivated by the occurrence of the site-specific operator $n_1$ in the numerator of the expression
of $c_\sV^\mu(q)$, we expand the expression for $H_\sV - \mu N_\sV$ to isolate the operators which apply to the site
labelled 1:
\begin{equation*}
	H_\sV - \mu N_\sV
	= (1-\mu)n_1 + \lambda n_1 (n_1 - 1) + \frac{n_1}{V}
		- \frac{a\astr_1}{V} \sum_{x\ne 1} a_x^\ast - \frac{a_1^\ast}{V} \sum_{x\ne 1} a\astr_x
		+ \widetilde{H}_\sV
\end{equation*}
where 
\[
	\widetilde{H}_\sV
=	(1-\mu)\sum_{x\ne1} n\astr_x - \frac{1}{V} \sum_{x,y\ne1}  a^\ast_x a\astr_y 
	+ \lambda \sum_{x\ne1} n_x ( n_x - 1)
\]
is an infinite-range-hopping Bose-Hubbard Hamiltonian for $\Lambda_\sV\setminus\{1\}$.
By denoting
\[
	h_{q,\sV} = (1 - \mu)(n_1 + q) + \lambda (n_1 + q) (n_1 + q - 1) +\tfrac{n_1}{V} 
\]
we may then write
\begin{equation*}
	\Hv{q}
	= h_{q,\sV} - \frac{a\astr_1}{V} \sum_{x\ne 1} a_x^\ast - \frac{a_1^\ast}{V} \sum_{x\ne 1} a\astr_x
		+ \widetilde{H}_\sV.
\end{equation*}
Note that $h_{q,\sV} \to h_q$ on $\mathcal{F}_{+}(\CC)$ as $V \to \infty$.

We intend to completely segregate the Hamiltonian $\Hv{q}$ into two individual parts, one which operates solely upon the 
site labelled 1, and the other which applies only to the remaining $V-1$ sites. What prevents us from doing this 
immediately is of course the  ``cross-term''
\[
	\frac{a\astr_1}{V} \sum_{x\ne 1} a_x^\ast + \frac{a_1^\ast}{V} \sum_{x\ne 1} a\astr_x.
\]
Motivated by the Approximating Hamiltonian technique, we shall substitute this term with
\[
	a\astr_1 \bar{R} + a_1^\ast R
\]
for a certain $c_0$-number $R$. Without loss of generality, we may take $R$ to be a 
non-negative real number. Fixing
\[
	h_{q,\sV}(R) = h_{q,\sV} - R ( a\astr_1 + a_1^\ast )
\]
then the resulting newly approximated Hamiltonian may be expressed as
\[
	\Happ{q}(R) = h_{q,\sV}(R) + \widetilde{H}_\sV.
\]
In the arguments that follow, we shall either take $R$ to equal $r_\mu$
in the variational principle, or a variable depending on $V$ which tends to $r_\mu$ in the limit.

\subsection{Case 1: \texorpdfstring{values of $\mu$ such that $r_\mu=0$ --} the absence of condensation}

First we shall state and prove the following:
\begin{proposition}								\label{PropB1}
For all $\mu\in\RR$ such that $r_\mu=0$,
\[
	\thermlim \frac{\trace_{\FFsym} \exp\big\{-\beta \Happ{q}(0) \big\}}
	{\trace_{\FFsym} \exp\big\{-\beta \Hv{q} \big\}} = 1.
\]
\end{proposition}

\begin{proof}
Using the Bogoliubov inequality:
\begin{equation} 								\label{bogoliubov}
	\la A-B \ra_B \le \ln \trace e^A - \ln \trace e^B \le \la A-B \ra_A
\end{equation}
for any $R$ we obtain
\begin{multline}								\label{sandwich}
	\beta \left\la a\astr_1 \left( \frac{\sum_{x\ne 1} a_x^\ast }{V} - R \right) 
	\right\ra_{\Happ{q}(R)} +\mathrm{h.c.}
\\	
\le	\ln \trace \exp\big\{-\beta \Hv{q} \big\} - \ln \trace \exp\big\{-\beta \Happ{q}(R) \big\}
\\	
\le	\beta \left\la a\astr_1 \left( \frac{\sum_{x\ne 1} a_x^\ast }{V} - R \right)
	\right\ra_{\Hv{q}} + \mathrm{h.c}.
\end{multline}
Taking the left-hand side, since $\Happ{q}(R)$ is a sum of two Hamiltonians which act upon different Hilbert spaces,
traces and therefore expectations may be easily de-coupled, so one may see that
\[
	\left\la a\astr_1 \left( \frac{\sum_{x\ne 1} a_x^\ast }{V}\right) \right\ra_{\Happ{q}(R)}
= 	\la a\astr_1 \ra_{h_{q,\sV}(R)}
	\left\la \frac{\sum_{x\ne 1} a_x^\ast}{V} \right\ra_{\widetilde{H}_\sV} = 0,
\]
which is zero since $\widetilde{H}_\sV$ is a gauge-invariant Hamiltonian:
$\left\la \sum_{x\ne 1} a_x^\ast \right\ra_{\widetilde{H}_\sV}=0$.

On the right-hand side, note that $\la a_1\ra_{\Hv{q}}=0$, also due to gauge invariance. Therefore we may simplify 
(\ref{sandwich}) to obtain
\begin{multline}							\label{sandwich1}
	-\beta R \big\la a\astr_1 + a^\ast_1 \big\ra_{h_{q,\sV}(R)}
\\
\le	\ln \trace \exp\big\{-\beta \Hv{q} \big\} - \ln \trace \exp\big\{-\beta \Happ{q}(R) \big\}
\\
\le	\beta \left\la a\astr_1 \left( \frac{\sum_{x\ne 1} a_x^\ast }{V}\right) \right\ra_{\Hv{q}} + \mathrm{h.c.}
\end{multline}

For this case we shall take $R=r_\mu = 0$. We therefore obtain
\begin{equation}\label{ineq}
0
\le	\ln \trace \exp\big\{-\beta \Hv{q} \big\} - \ln \trace \exp\big\{-\beta \Happ{q}(0) \big\}
\le	\beta \left\la a\astr_1 \left( \frac{\sum_{x\ne 1} a_x^\ast }{V}\right) \right\ra_{\Hv{q}} + \mathrm{h.c.}
\end{equation}
Now by the Schwarz inequality
\begin{align*}
	\left| \left\la a\astr_1 \left( \frac{\sum_{x\ne 1} a_x^\ast }{V} \right) \right\ra_{\Hv{q}} \right|
&=	\left| \left\la a\astr_1 \left( \frac{\sum_{x} a_x^\ast }{V}\right) - \frac{a\astr_1 a_1^\ast}{V} \right\ra_{\Hv{q}} \right|
\\
&\le	\la  n_1 \ra_{\Hv{q}}^{\tfrac{1}{2}} \left\la \frac{\sum a_x^\ast \sum a\astr_x}{V^2}\right\ra^{\tfrac{1}{2}}_{\Hv{q}}
	+ \frac{\la n_1 \ra_{\Hv{q}}+1}{V}.
\end{align*}

To consider this let $H^s_{q,\sV} \vcentcolon= \Hv{q} + s\frac{\sum a_x^\ast \sum a\astr_x}{V}$
and $\widehat{H}^s_{q,\sV} \vcentcolon= H^s_{q,\sV} - 2 \lambda q n_1$. Using the Bogoliubov inequality
(\ref{bogoliubov}), with $A = -\beta \widehat{H}^s_{q,\sV}$ and $B = -\beta H^s_{q,\sV}$, so that 
$A-B = 2 \beta \lambda q n_1$, one has
\begin{equation}										\label{upperbound}
	2 \beta \lambda q \la n_1 \ra_{H^s_{q,\sV}} 
\le	\ln \trace \e^{-\beta \widehat{H}^s_{q,\sV}} - \ln \trace \e^{-\beta H^s_{q,\sV}}
\le 	2 \beta \lambda q \la n_1 \ra_{\widehat{H}^s_{q,\sV}} 
=	2 \beta \lambda q \frac{ \la N \ra_{\widehat{H}^s_{q,\sV}} }{V}.
\end{equation}
The last equality is due to the fact that the system $\widehat{H}^s_{q,\sV}$ is invariant under
permutation of the sites of the lattice. This identity implies the following:
\begin{enumerate}
\renewcommand{\labelenumi}{(\roman{enumi})}
\item With $s=0$ we get $\la n_1 \ra_{\Hv{q}} \le \la N \ra_{H_\sV} /V$, thus $\la n_1 \ra_{\Hv{q}}$ is bounded and 
	in the limit $\frac{\la n_1 \ra_{\Hv{q}}}{V} \to 0$.
\item \[
		0 \le \frac{1}{V} \ln \trace \e^{-\beta \widehat{H}^s_{q,\sV}} - \frac{1}{V} \ln \trace \e^{-\beta H^s_{q,\sV}}
		\le \frac{2 \beta \lambda q}{V} \frac{ \la N \ra_{\widehat{H}^s_{q,\sV}} }{V}
	\] indicating that in the limit, the pressures are the same for the two Hamiltonians. By using Griffith's Lemma 
	we see that the condensate densities (the derivatives with respect to $s$ at zero) are both equal to zero 
	(since we are considering the case $r_\mu=0$ here).
	That is
	\[
		\thermlim \left\la \frac{\sum a_x^\ast \sum a\astr_x}{V^2}\right\ra_{\Hv{q}}
		= \thermlim \left\la \frac{\sum a_x^\ast \sum a\astr_x}{V^2}\right\ra_{H_\sV}
		\vcentcolon= 0.
	\]
\end{enumerate}
Using these facts, one sees that the right-hand side of (\ref{ineq}) goes to zero in the limit
and we can conclude that
\[
	\thermlim \frac{\trace_{\FFsym} \exp\big\{- \beta \Hv{q}\big\}}
		{\trace_{\FFsym} \exp\big\{-\beta \Happ{q}(0)\big\}}=1.
\]
\end{proof}

Using Proposition \ref{PropB1}, one immediately obtains the desired result:
\begin{align*}
	c^\mu_{\phantom{l}}(q)
&=	\thermlim c^\mu_\sV(q) = \thermlim \frac{\trace_{\FFsym} 
	\exp\big\{- \beta \Hv{q}\big\}}{\trace_{\FFsym} \exp\big\{-\beta \Hv{0}\big\}}
\\
&=	\thermlim \frac{\trace_{\FFsym} 
	\exp\big\{- \beta \Happ{q}(0)\big\}}{\trace_{\FFsym} \exp\big\{-\beta \Happ{0}(0)\big\}}
\\
& =	\thermlim \frac{\trace_{\mathcal{F}_+(\mathcal{H}_{\sV-1})} 
	\exp\big\{- \beta \widetilde{H}_\sV  \big\}
	}
	{\trace_{\mathcal{F}_+(\mathcal{H}_{\sV-1})} 
	\exp\big\{- \beta \widetilde{H}_\sV \big\}
	}
	\frac{\trace_{\mathcal{F}(\CC)} \exp\big\{- \beta h_{q,\sV} \big\}}
	{\trace_{\mathcal{F}(\CC)} \exp\big\{-\beta h_{0,\sV} \big\}} 
\\
& = 	\frac{\trace_{\mathcal{F}(\CC)} 
	\exp\big\{- \beta h_{q}(0) \big\}}
	{\trace_{\mathcal{F}(\CC)} \exp\big\{-\beta h_{0}(0) \big\}}
\end{align*} 
in the limit since $h_{q,\sV} = h_{q,\sV}(0) \to h_q(0)$ on $\mathcal{F}(\CC)$.

\subsection{Case 2: For any \texorpdfstring{$\mu$}{chemical potential}}
Considering the case when $r_\mu > 0$, if we insert $R = r_\mu$ in the constraint inequality (\ref{sandwich1})
then its left-hand term is strictly negative, but its right-hand term is strictly positive, rendering the 
previous argument useless here.

We therefore introduce a gauge-breaking term $\bar{\nu} \sum_x a\astr_x+ \nu \sum_x a^\ast_x$ into
the Hamiltonians $\Hv{q}$ and $\Happ{q}(r_\mu)$. Without loss of generality we may assume $\nu$ to 
be real and positive, so denote
\[
	\Hv{q}(\nu) = \Hv{q} - \nu \sum_{x=1}^V ( a\astr_x + a_x^\ast)
\]
and its corresponding approximation as
\[
	\Happ{q}(R,\nu) = \Happ{q}(R) - \nu \sum_{x=1}^V ( a\astr_x + a_x^\ast).
\]
Again we wish to separate this Hamiltonian into parts, one acting upon the site labelled 1, the
other on the remaining $V-1$ sites. If we define:
\begin{align*}
	h_{q,\sV}(r,\nu) 
&\vcentcolon=	h_{q,\sV}(r) - \nu (a_1 + a^\ast_1)
\\	
&=	(1 - \mu)(n_1 + q) + \lambda (n_1 + q) (n_1 + q - 1) +\tfrac{n_1}{V} - (r+\nu)(a_1 + a^\ast_1).
\end{align*}
and
\[
	\widetilde{H}_\sV(\nu) = \widetilde{H}_\sV - \nu \sum_{x\ne1} (a\astr_x + a^\ast_x)
\]
then we may write $\Happ{q}(R,\nu) = h_{q,\sV}(R,\nu) + \widetilde{H}_\sV(\nu)$. 
Denote $\thermlim h_{q,\sV}(r,\nu) = h_{q}(r,\nu)$ on $\mathcal{F}_{+}(\CC)$, i.e. obtain a 
``gauge symmetry broken'' single site Hamiltonian
\[
	h_{q}(r,\nu) = (1 - \mu)(n + q) + \lambda (n + q) (n + q - 1) - (r+\nu)(a + a^\ast).
\]
We shall first prove the following proposition:

\begin{proposition}									\label{PropB2}
For each $\nu>0$, there exists a sequence $\{ \nu_\sV \in \RR: \nu_\sV \in [\nu, \nu+1/\sqrt{V}]\}$
independent of $q$ such that
\begin{equation*}
	\thermlim \frac{\trace_{\FFsym} \exp\left\{-\beta \Hv{q}(\nu_\sV)\right\}}
	{\trace_{\FFsym} \exp\left\{-\beta \Happ{q}(r_\mu(\nu_\sV), \nu_\sV)\right\}} = 1
\end{equation*}
where $r_\mu(\nu)>0$ is the non-zero solution of
\begin{equation}									\label{r-nu-value}
	2 r
=	\la a + a^\ast \ra_{h_q(r,\nu) }.
\end{equation}
Note that $\lim_{\nu\to0} r_\mu(\nu) = r_\mu$, the maximal solution of $(\ref{euler-lagrange})$,
i.e. the positive square root of the condensate density. 
\end{proposition}

\begin{proof}
There is no immediate correlation between the chosen $R$ and $r_\mu$ as yet. For each $\nu>0$
take a sequence $\nu_\sV$ which tends to $\nu$ as $V \to \infty$.
Then using the Bogoliubov inequality again, we obtain
\begin{multline}					\label{gauge_invar_bogoulibov}
	\beta \left\la a\astr_1 \left( \frac{\sum_{x\ne 1} a_x^\ast }{V}
	- R\right) \right\ra_{\Happ{q}(R,\,\nu_\sV)} + \mathrm{h.c.}
\\
	\le
	\ln \trace \exp\left\{-\beta \Hv{q}(\nu_\sV) \right\} - \ln \trace \exp\left\{-\beta \Happ{q}(R,\nu_\sV)\right\}
\\
	\le
	\beta \left\la a\astr_1 \left( \frac{\sum_{x\ne 1} a_x^\ast }{V} 
	- R\right) \right\ra_{\Hv{q}(\nu_\sV)} + \mathrm{h.c}.
\end{multline}

As above, the left-hand side may be reduced to
\[
	\left\la a\astr_1 \left( \frac{\sum_{x\ne 1} a_x^\ast }{V}-R \right) \right\ra_{\Happ{q}(R,\,\nu_\sV)}
   = 	\la a\astr_1 \ra_{h_{q,\sV}(R,\nu_\sV)}
	\left\la \frac{\sum_{x\ne 1} a_x^\ast}{V} - R
	\right\ra_{\widetilde{H}_\sV(\nu_\sV)}.
\]
If we replace $R$ with the term
\begin{equation}								\label{r_choice}
	r^{-}_{\mu,\sV}(\nu_\sV) = \left\la \frac{\sum_{x\ne 1} a_x}{V} \right\ra_{\widetilde{H}_\sV(\nu_\sV)},
\end{equation}
then the left-most side of (\ref{gauge_invar_bogoulibov}) is identically zero and we get that
\[
	0
\le	\ln \trace \exp\left\{-\beta \Hv{q}(\nu_\sV)\right\} - 
	\ln \trace \exp\left\{-\beta \Happ{q}(r^{-}_{\mu,\sV}(\nu_\sV),\nu_\sV)\right\}.
\]
Hence
\begin{equation}								\label{liminf-ineq}
	\liminf_{V\to\infty} \frac{\trace \exp\left\{-\beta \Hv{q}(\nu_\sV)\right\}}
	{\trace \exp\left\{-\beta \Happ{q}(r^{-}_{\mu,_\sV}(\nu_\sV), \nu\sV)\right\}} \ge 1.
\end{equation}
\\

Now considering the right-hand side of (\ref{gauge_invar_bogoulibov}) for any $\mu$. 
Using the Schwarz inequality as before:
\begin{align}
&	\left| \left\la a\astr_1 \left( \frac{\sum_{x\ne 1} a_x^\ast }{V}
	- R \right) \right\ra_{\Hv{q}(\nu)} \right|
=	\left| \left\la a\astr_1 \left( \frac{\sum_{x} a_x^\ast }{V} - R \right)
	- \frac{a\astr_1 a_1^\ast}{V} \right\ra_{\Hv{q}(\nu)} \right|		\notag
\\[0.15cm]
&\qquad\le	\la n_1 \ra_{\Hv{q}(\nu)}^{\tfrac{1}{2}}
	\left\la \left( \frac{\sum_{x} a_x^\ast }{V} - R \right)\left( \frac{\sum_{x} a\astr_x }{V} - R \right)  
	\right\ra^{\tfrac{1}{2}}_{\Hv{q}(\nu)} + \frac{\la n_1 \ra_{\Hv{q}(\nu)}+1}{V}	\notag
\\[0.3cm]
&\qquad=	\left( \la n_1 \ra_{\Hv{q}(\nu)} \frac{1}{V} 
	\big\la \delta_0^\ast \delta\astr_0 \big\ra_{\Hv{q}(\nu)} \right)^{1/2} 
	+ \frac{\la n_1 \ra_{\Hv{q}(\nu)}+1}{V}					\label{ineq2}
\end{align}
where we have taken
\[	
	\delta\astr_0 = \frac{1}{\sqrt{V}} \left( \sum_{x=1}^V a\astr_x - V R \right).
\]

Again in order to consider this, insert
$\Hv{q}(\nu)$ and $\widehat{H}_{q,\sV}(\nu) \vcentcolon= \Hv{q}(\nu) - 2 \lambda q n_1$ 
into the Bogoliubov inequality, to obtain
\begin{equation}									\label{bog-5}
	2 \beta \lambda q \la n_1 \ra_{\Hv{q}(\nu)} 
\le	\ln \trace \e^{-\beta \widehat{H}_{q,\sV}(\nu)} - \ln \trace \e^{-\beta \Hv{q}(\nu)}
\le	2 \beta \lambda q \la n_1 \ra_{\widehat{H}_{q,\sV}(\nu)} 
=	2 \beta \lambda q \frac{ \la N \ra_{\widehat{H}_{q,\sV}(\nu)} }{V}
\end{equation}
which implies the following facts:
\begin{enumerate}
\renewcommand{\labelenumi}{(\roman{enumi})}
\item $\la n_1 \ra_{\Hv{q}(\nu)} \le \la N \ra_{\widehat{H}_{q,\sV}(\nu)}/V$, and 
	hence $\frac{\la n_1 \ra_{\Hv{q}(\nu)}}{V} \to 0$ as $V\to\infty$.
\item Since $\widehat{H}_{q,\sV}(\nu) = (q-\mu)q + H_\sV(\nu) - \mu N_\sV$,
	\[
		0 \le -\frac{ \beta(q-\mu)q }{V} + \frac{1}{V} \ln \trace \e^{-\beta ( H_\sV(\nu) - \mu N_\sV )} 
		- \frac{1}{V} \ln \trace \e^{-\beta \Hv{q}(\nu)}
		\le \frac{2 \beta \lambda q}{V} \frac{ \la N \ra_{H_\sV(\nu)} }{V}
	\] with which one may show that in the limit, the pressures are the same for $H_\sV(\nu)$ 
	and $\Hv{q}(\nu)$:
	\begin{equation}							\label{Pressures2}
		\thermlim p_\sV[ H_\sV(\nu) ] = \thermlim p_\sV[ \Hv{q}(\nu) ].
	\end{equation}
\end{enumerate}

Using fact (i) from above, we find that the only term we need yet be concerned on the right hand side of 
(\ref{gauge_invar_bogoulibov}) is the first term of (\ref{ineq2}), whose behaviour in the thermodynamic 
limit is still unknown:
\[
	\frac{1}{V} \big\la \delta_0^\ast \delta\astr_0 \big\ra_{\Hv{q}(\nu)}.
\]
To deal with this we shall take $R$ to be the following:
\begin{equation}								\label{RHS_r_val}
	r^{+}_{\mu,\sV}(\nu) = \frac{1}{V} \left\la \sum_{x=1}^V a_x \right\ra_{\Hv{q}(\nu)}
\end{equation}
so that one has
\begin{equation}								\label{delta0}
	\delta_0 = \frac{1}{\sqrt{V}} \left( \sum_{x=1}^V a_x - \bigg\la \sum_{x=1}^V a_x \bigg\ra_{\Hv{q}(\nu)} \right).
\end{equation}
Now we shall state and use some lemmas, which are proved later: 

\begin{lemma} 								\label{lemma1}
For fixed $\nu>0$, a positive integer $q$, and $\delta_0$ is defined as \textnormal{(\ref{delta0})},
then there exists a sequence $\{ \nu_\sV \in \RR : \nu_\sV \in [\nu, \nu+1/\sqrt{V}] \}$ independent of $q$
which tends to $\nu$ as $V \to \infty$, such that for large $V$ we have the approximation
\begin{equation*}
	\left\la \delta_0^\ast \delta\astr_0 \right\ra_{\Hv{q}(\nu_\sV)}
\le	B \e^{q} \left(2\sqrt{V} + \frac{1}{\nu}\right) + 
	u\frac{q}{V} + w
\end{equation*}
for some constants $u$, $w$ and $B$, independent of $\nu$ and $q$.
\end{lemma}

Using this lemma, for large $V$ and fixed $q$, we obtain the following estimate
\begin{equation*}
	\ln \trace \exp\left\{-\beta \Hv{q}(\nu_\sV)\right\} - 
	\ln \trace \exp\left\{-\beta \Happ{q}(r^{+}_{\mu,\sV}(\nu_\sV),\nu_\sV)\right\}
	\le \frac{\text{const}}{V^{1/4}}
\end{equation*}
where $\nu_\sV \to \nu$ as $V \to \infty$, implying that 
\begin{equation}							\label{limsup-ineq}
	\limsup_{V\to\infty} \frac{\trace \exp\left\{-\beta \Hv{q}(\nu_\sV)\right\}}
	{\trace \exp\left\{-\beta \Happ{q}(r^{+}_{\mu,\sV}(\nu_\sV),\nu_\sV)\right\}} \le 1.
\end{equation}

\begin{lemma}							\label{lemma3-r_vals-equal}
For a fixed $\nu>0$ and for any sequence $\{ \nu_\sV \}$ which tends to $\nu$ as $V \to \infty$, then
\begin{equation*}
	\thermlim r^{-}_{\mu,\sV}( \nu_\sV) 
=	\thermlim r^{+}_{\mu,\sV}( \nu_\sV)
= 	r_\mu(\nu)
\end{equation*}
where $r_\mu(\nu)$ is the unique non-zero solution to the Euler-Lagrange equation $(\ref{r-nu-value})$.
\end{lemma}

For clarity, it is best to use the following short-hand for this argument:
\begin{align*}
	a_\sV &= \trace_{\mathcal{F}(\Hone)} \exp\big\{-\beta \Hv{q}(\nu_\sV) \big\}
\\
	b_\sV &= \trace_{\mathcal{F}(\Hone)} \exp\big\{-\beta \Happ{q}(r_{\mu}(\nu_\sV),\nu_\sV)\big\}
\\
	c_\sV &= \trace_{\mathcal{F}(\Hone)} \exp\big\{-\beta \Happ{q}(r^{-}_{\mu,\sV}(\nu_\sV),\nu_\sV)\big\}
\\
	d_\sV &= \trace_{\mathcal{F}(\Hone)} \exp\big\{-\beta \Happ{q}(r^{+}_{\mu,\sV}(\nu_\sV),\nu_\sV)\big\}
\end{align*}

The penultimate step is to prove the following:
\begin{equation*}
	\thermlim \frac{b_\sV}{c_\sV} = 1 \qquad \quad \text{and} \quad \qquad
	\thermlim \frac{b_\sV}{d_\sV} = 1.
\end{equation*}
Considering the first, note that
\[
	\frac{b_\sV}{c_\sV} 
=	\frac{\trace_{\mathcal{F}(\CC)} \e^{-\beta h_q(r_{\mu}(\nu_\sV), \nu_\sV)}}
	{\trace_{\mathcal{F}(\CC)} \e^{-\beta h_q(r^{-}_{\mu,\sV}(\nu_\sV), \nu_\sV)}}.
\]
Once again by the Bogoliubov inequality (\ref{bogoliubov}), with $A=-\beta h_q(r^{-}_{\mu,\sV}(\nu_\sV), \nu_\sV)$ and 
$B=-\beta h_q(r_{\mu}(\nu_\sV), \nu_\sV)$
then $A-B = \beta (r_\mu(\nu_\sV) - r^{-}_{\mu, \sV}(\nu_\sV) )( a + a^\ast )$ and we obtain
\begin{multline}								\label{sandwich2}
	\beta (r_\mu(\nu_\sV) - r^{-}_{\mu, \sV}(\nu_\sV) ) 
	\left\la a + a^\ast \right\ra_{h_q(r_{\mu}(\nu_\sV), \nu_\sV)}
\\
\le	\ln b_\sV  - \ln c_\sV
\le 	\beta (r_\mu(\nu_\sV) - r^{-}_{\mu, \sV}(\nu_\sV) ) 
	\left\la a + a^\ast \right\ra_{h_q(r^{-}_{\mu, \sV}(\nu_\sV), \nu_\sV)}.
\end{multline}
By continuity one may see that $\thermlim r_\mu(\nu_\sV) = r_\mu(\nu)$. 
Then as $V \to \infty$, both the left and right hand sides of (\ref{sandwich2}) go to zero, implying the first result. 
A similar procedure may be used to show the second.

We want to show that $\thermlim \tfrac{a_\sV}{b_\sV} = 1$. Since (\ref{limsup-ineq}) implies 
that $\limsup \tfrac{a_\sV}{d_\sV} \le 1$, we have:
\[
	\limsup_{V \to \infty} \frac{a_\sV}{b_\sV}
=	\limsup_{V \to \infty} \frac{\frac{a_\sV}{d_\sV}}{\frac{b_\sV}{d_\sV}}
\le 	 \frac{ \limsup_{V \to \infty} \frac{a_\sV}{d_\sV}}{ \thermlim \frac{b_\sV}{d_\sV}}
\le 	1.
\]
The infimum limit follows similarly from (\ref{liminf-ineq}), proving the proposition.
\end{proof}

With the assistance of Proposition \ref{PropB2} we then have our result:
\begin{align*}
	&\thermlim c^\mu_\sV(q, \nu_\sV)
=	\thermlim \frac{\trace_{\FFsym} \exp\big\{-\beta \Hv{q}(\nu_\sV)\big\}
	}{
	\trace_{\FFsym} \exp\big\{-\beta \Hv{0}(\nu_\sV)\big\}
	}
\\
&=	\thermlim \frac{\trace_{\FFsym} \exp\big\{-\beta \Happ{q}(r_{\mu,\sV}(\nu_\sV),\nu_\sV)\big\}
	}{
	\trace_{\FFsym} \exp\big\{-\beta \Happ{0}(r_{\mu,\sV}(\nu_\sV),\nu_\sV)\big\}
	}
\\
&=	\thermlim \frac{\trace_{\mathcal{F}_{+}(\mathcal{H}_{\sV-1})}
		\exp\big\{-\beta \widetilde{H}_\sV(\nu_\sV) \big\}
	}{
		\trace_{\mathcal{F}_{+}(\mathcal{H}_{\sV-1})} 
			\exp\big\{-\beta \widetilde{H}_\sV(\nu_\sV)\big\}
	}
	\frac{\trace_{\mathcal{F}(\CC)}
		\exp\big\{-\beta  h_{q,\sV}(r_{\mu,\sV}(\nu_\sV), \nu_\sV) \big\}
	}{
		\trace_{\mathcal{F}(\CC)} 
		\exp\big\{-\beta  h_{0,\sV}(r_{\mu,\sV}(\nu_\sV), \nu_\sV)\big\}
	}
\\
&=	\frac{\trace_{\mathcal{F}(\CC)}
		\exp\big\{-\beta [h_q(r_\mu(\nu)) - \nu(a + a^\ast)]\big\}
	}{
		\trace_{\mathcal{F}(\CC)}
		\exp\big\{-\beta [h_0(r_\mu(\nu)) - \nu(a + a^\ast)]\big\}
	}.
\end{align*}

\subsection{Proof of Lemma \ref{lemma1}}
\begin{proof}  
In the Appendix we prove that (see (\ref{bound-8})\!) there exist constants $u$ and $w$ such that
for fixed $\nu_0 \in \RR$, all $\nu < \nu_0$ and all $q \in \NN$,
\begin{equation}										\label{bound-7}
	\left\la \delta_0^\ast \delta\astr_0 \right\ra_{\Hv{q}(\nu)}
\le	f_{q,\sV}(\nu) + u\frac{q}{V} + w
\end{equation}
with $f_{q,\sV}(\nu) \vcentcolon= (\delta_0, \delta_0 )_{\Hv{q}(\nu)}$ where $(\,\cdot\,,\,\cdot\,)_H$ is the 
Duhamel inner product, see (\ref{bog-inner-prod}). Set $c_0 = \tfrac{1}{\sqrt{V}} \sum a_x$.

One may check that for $\nu > 0$:
\begin{equation}									\label{second-deriv-polar}
	f_{q,\sV}(\nu)
	= \frac{1}{4\beta\nu} \frac{\partial}{\partial \nu} 
	\left( \nu \frac{\partial}{\partial \nu} p_\sV[\Hv{q}(\nu)] \right).
\end{equation}
This follows from the fact that when $\nu \in \CC$, one can write $f_{q,\sV}(\nu) =\partial_\nu 
\partial_{\bar{\nu}} p_\sV[\Hv{q}(\nu)]$,
but since $p_\sV[\Hv{q}(\nu)]$ does not depend on the argument of $\nu$
we can use polar coordinates to get (\ref{second-deriv-polar}).
We want to show that $\thermlim f_{q,\sV}(\nu)/V \to 0$.
Consider (\ref{second-deriv-polar}), multiply both sides by $\nu$ and integrate:
\[
	\int_\nu^{\nu+1/\sqrt{V}} \nu' f_{q,\sV}(\nu') d\nu' = \frac{1}{4\beta}  
	\left( \nu' \frac{\partial}{\partial \nu'} p_\sV[\Hv{q}(\nu')] \right) \bigg|_\nu^{\nu+1/\sqrt{V}} 
\]
for $[\nu,\nu+1/\sqrt{V}] \subset [0, \nu_0]$. 
Using the bound (\ref{firstderiv-bound}) proved in the Appendix, then there is a constant B such that
for all $q \in \NN$ we obtain
\begin{equation}									\label{int-f-bound}
	\int_\nu^{\nu+1/\sqrt{V}} \nu' f_{q,\sV}(\nu') d\nu' 
\le 	B \left(2\nu+\frac{1}{\sqrt{V}} \right) .
\end{equation}
Now let
\begin{equation}									\label{Fseries}
	F_\sV(\nu) = \sum_{q=1}^\infty \e^{-q}  f_{q,\sV} (\nu).
\end{equation}
In the Appendix (see (\ref{second-deriv-line1})--(\ref{f-bound-bad})\!) we show that there exist constants $a$ and $b$ independent of $q$
such that for all $\nu<\nu_0$ and $q\in\NN$:
\[
	f_{q,\sV}(\nu) \le \frac{a}{\nu} +  b V.
\]
Therefore the series (\ref{Fseries}) is uniformly convergent in $\nu$ and since each term 
is continuous in $\nu$, $F_\sV(\nu)$ is also continuous. From (\ref{int-f-bound}) we obtain:
\[
	\int_\nu^{\nu+1/\sqrt{V}} \nu' F_\sV(\nu') d\nu' 
\le 	\frac{B}{\e-1} \left(2\nu+\frac{1}{\sqrt{V}}\right)
\le 	B \left(2\nu+\frac{1}{\sqrt{V}}\right).
\]
By the Mean-Value theorem, there exists some $\nu_\sV \in [\nu, \nu+1/\sqrt{V}]$ (independent of $q$) such that
\begin{equation}									\label{mvt}
	\int_\nu^{\nu+1/\sqrt{V}} \nu' F_\sV(\nu') d\nu' 
=	\frac{\nu_\sV}{\sqrt{V}} F_\sV(\nu_\sV)
\end{equation}
which implies that
\[
	F_\sV(\nu_\sV)
\le	B \left( 2\sqrt{V} \frac{\nu}{\nu_\sV} +\frac{1}{\nu_\sV} \right).
\]
For any positive integer $q$, since $\e^{-q} f_{q,\sV}(\nu) \le F_\sV(\nu)$, then
\begin{align*}
	f_{q,\sV}(\nu_\sV)
&\le 	B \e^{q} \left( 2\sqrt{V} \frac{\nu}{\nu_\sV} +\frac{1}{\nu_\sV} \right).
\end{align*}
Thus we have a sequence 
$\{ \nu_\sV  \in \RR: \nu_\sV \in [\nu, \nu+1/\sqrt{V}] \}$ satisfying (\ref{mvt}), independent of $q$,
which tends to $\nu>0$ as $V \to \infty$, such that for large $V$ we have the estimate:
\[
	f_{q,\sV}(\nu) \le B \e^q \left( 2\sqrt{V}  + \frac{1}{\nu}\right).
\]
Combine this with (\ref{bound-7}) to complete the proof.
\end{proof}

\subsection{Proof of Lemma \ref{lemma3-r_vals-equal}}
\begin{proof}
Before proceeding, we need to show the following: for fixed $\nu >0$
\[
	\thermlim p_\sV[ \widetilde{H}_\sV(\nu) ] = \thermlim p_\sV[ H_\sV(\nu) ],
\]
i.e. a single site's contribution is irrelevant in the thermodynamic limit.
Recall that
\begin{align*}
	\widetilde{H}_\sV
&=	\frac{1}{2V} \sum_{x,y\ne1} ( a^\ast_x - a^\ast_y )( a\astr_x - a\astr_y ) 
	+ \lambda \sum_{x\ne1} n_x ( n_x - 1)
	+ \left(\frac{2}{V(V-1)} - \mu \right) \sum_{x\ne1} n_x.
\end{align*}
The corresponding pressure may be expressed as
\begin{align*}
	p_\sV[ \widetilde{H}_\sV ]
&=	\frac{1}{\beta V} \ln \trace_{\FFsym} \exp\left\{
	-\beta \widetilde{H}_\sV
	\right\}
\\
&=	\frac{1}{\beta V} \ln \trace_{\mathcal{F}_{+}(\mathcal{H}_{V-1})} \exp\bigg\{
	-\beta \bigg( \frac{1}{2V} \sum_{x,y=1}^{V-1} ( a^\ast_x - a^\ast_y )( a\astr_x - a\astr_y ) 
	+ \lambda \sum_{x=1}^{V-1} n_x ( n_x - 1)
\\
&	\qquad \qquad  + \left(\frac{2}{V(V-1)} - \mu \right) \sum_{x=1}^{V-1} n_x 
	\bigg)
	\bigg\}
\\
&=	\left( \frac{V-1}{V} \right)^2
	p_{\sV-1}\big[ H_{\sV-1}( \beta\left( \tfrac{V-1}{V} \right) , \lambda\left( \tfrac{V}{V-1} \right),
	\mu - \tfrac{2}{V(V-1)} )\big]
\end{align*}
where abusing notation temporarily we have explicitly included the parameters of the 
IRH Bose-Hubbard Hamiltonian,
i.e. we write the pressure of (\ref{I-R}) as $p_\sV[ H_\sV ] \equiv p_\sV[ H_\sV(\beta,\lambda,\mu)]$.
Then in the limit, with the use of the Bogoliubov inequality, one may verify that
\begin{align*}
	\thermlim p_\sV[\widetilde{H}_\sV]
&=	\thermlim \left( \frac{V-1}{V} \right)^2
	p_{\sV-1}\big[ H_{\sV-1}( \beta\left( \tfrac{V-1}{V} \right) , \lambda\left( \tfrac{V}{V-1} \right),
	\mu - \tfrac{2}{V(V-1)} )\big]
\\
&=	\thermlim p_\sV[H_\sV ] \equiv p(\beta,\mu).
\end{align*}

Now proceeding to prove this lemma, recall that we chose
\[
	r^{-}_{\mu,\sV}(\nu) = \left\la \frac{\sum_{x\ne 1} a_x}{V} \right\ra_{\widetilde{H}_\sV(\nu)}
\]
where $\widetilde{H}_\sV(\nu) = \widetilde{H}_\sV - \nu \sum_{x\ne1} ( a\astr_x + a_x^\ast )$
is the gauge-broken IRH Bose-Hubbard Hamiltonian on
all sites of the lattice barring the site $x=1$. Fixing a value of $\nu>0$,
there exists a unique $r_\mu(\nu)>0$ as the solution to the Euler-Lagrange equation (\ref{r-nu-value}), i.e.
\[
	2r_\mu(\nu) = \la a + a^\ast \ra_{h_q(r_\mu(\nu),\nu)}.
\]
The pressure $p_\sV[ \widetilde{H}_\sV(\nu) ]$ is convex in $\nu$ and its thermodynamic limit 
is differentiable for all $\nu > 0$. By Griffith's Lemma, we have
\begin{equation}								\label{griff1}
	\thermlim \frac{d}{d\nu} p_\sV[ \widetilde{H}_\sV( \nu_\sV) ]
=	\frac{d}{d\nu} \thermlim p_\sV[ \widetilde{H}_\sV(\nu) ].
\end{equation}

The left hand side of this evaluates to
\[
	\thermlim \frac{d}{d\nu_\sV} p_\sV[ \widetilde{H}_\sV( \nu_\sV) ]
=	\thermlim \frac{1}{V} \sum_{x\ne1} \la a_x + a^\ast_x \ra_{\widetilde{H}_\sV(\nu_\sV)}
=	\thermlim \frac{2}{V} \sum_{x\ne1} \la a_x \ra_{\widetilde{H}_\sV(\nu_\sV)}
=	2 \thermlim r^{-}_{\mu,\sV}(\nu_\sV).
\]
As shown above, we have that
\begin{multline}								\label{pressures2}
	\thermlim p_\sV[ \widetilde{H}_\sV(\nu) ]
=	\thermlim p_\sV [ H_\sV(\nu) ]
\\
=	-r_\mu(\nu)^2 + \frac{1}{\beta} \ln \trace \exp\big\{ \beta 
	\big[ (\mu-1) n -\lambda n(n-1) + (r_\mu(\nu)+\nu)(a+a^\ast)\big]\big\}.
\end{multline}
so the right-hand side of (\ref{griff1}) will become (also using (\ref{pressures2})\!)
\begin{align*}
	\frac{d}{d\nu} & \thermlim p_\sV [ H_\sV(\nu) ]
\\
&=	\frac{d}{d\nu} \left[-r_\mu(\nu)^2 + \frac{1}{\beta} \ln \trace \exp\big( \beta 
	\big[ (\mu-1) n -\lambda n(n-1) + (r_\mu(\nu)+\nu)(a+a^\ast)\big]\big) \right]
\\
&=	-2r_\mu(\nu) \frac{dr_\mu(\nu)}{d\nu} + \left( \frac{dr_\mu(\nu)}{d\nu} + 1 \right) \la a 
	+ a^\ast \ra_{h_q(r_\mu(\nu),\nu)}
\\
&=	\frac{dr_\mu(\nu)}{d\nu} \left( \la a + a^\ast \ra_{H(\nu)} - 2r_\mu(\nu) \right) + \la a 
	+ a^\ast \ra_{h_q(r_\mu(\nu),\nu)}
\\
&=	2r_\mu(\nu).
\end{align*}
as desired.

Similarly, taking
\[
	r_{\mu,\sV}^{+}(\nu) = \frac{1}{V} \bigg\la \sum_{x=1}^V a_x \bigg\ra_{\Hv{q}(\nu)}
\]
where $\Hv{q}(\nu) = \Hv{q} - \nu \sum_{x=1}^V ( a\astr_x + a^\ast_x )$.
Label the corresponding pressure for this Hamiltonian as $p_\sV[ \Hv{q}(\nu)]$. Recall the expression 
(\ref{Pressures2}) that we previously derived: 
\[
	\thermlim p_\sV[ \Hv{q}(\nu)] = \thermlim p_\sV[ H_\sV(\nu)].
\]
As above by Griffith's Lemma, we have
\[
	\thermlim r^{+}_\sV(\nu_\sV) 
= 	\frac{1}{2} \thermlim \frac{d}{d\nu_\sV} p_\sV[ \Hv{q}( \nu_\sV) ]
=	\frac{1}{2} \frac{d}{d\nu} \thermlim p_\sV[ H_\sV(\nu) ]
=	r_\mu(\nu)
\]
as above.
\end{proof}

\section{Proof of Theorem \ref{no-long}}
For those values of $\mu$ such that $r_\mu = 0$, i.e. in the absence of condensation,
first note that the density (from (\ref{pressure})) may be expressed as:
\begin{multline*}
	\rhodens
= 	\frac{\partial}{\partial \mu} p(\beta,\mu)
=	\frac{\partial}{\partial \mu} 
	\frac{1}{\beta}\trace_{\mathcal{F}(\CC)} \exp \big\{ -\beta [
		 (1 - \mu)n +\lambda n(n - 1)
	] \big\}
\\
=	\frac{\trace_{\mathcal{F}(\CC)} \big[ n \exp \big\{ -\beta [
		(1 - \mu)n +\lambda n(n - 1)
	] \big\} \big]}
	{\trace_{\mathcal{F}(\CC)} \exp \big\{ -\beta [
		(1 - \mu)n +\lambda n(n - 1)
	] \big\}}.
\end{multline*}
Similarly, from Theorem \ref{prop3-equiv} one immediately obtains:
\[
	c^\mu(q,0)
   =	\frac{ \trace_{\mathcal{F}(\CC)} \exp \big\{ -\beta [
   		(1 - \mu)(n + q) + \lambda (n + q)(n + q - 1)
	] \big\} }
	{ \trace_{\mathcal{F}(\CC)} \exp \big\{ -\beta [
		(1 - \mu)n +\lambda n(n - 1)
	] \big\} }.
\]
Label the denominator $\Xi \vcentcolon= \trace_{\mathcal{F}(\CC)} \exp\{-\beta ( (1 - \mu)n + \lambda n(n - 1) )\}$.
The operator $n$ in this context counts the number of particles on the site, so in terms of a basis of 
occupation numbers, it has eigenvalues $k=0,1,2,\dots$. Summing over this basis
\begin{align*}
	\sum_{q=1}^\infty c^\mu(q,0)
   =&	\frac{1}{\Xi} \sum_{q=1}^\infty \sum_{k=0}^\infty \exp \big\{ -\beta [
		\lambda (1 - \mu)(k + q) + \lambda (k + q)(k + q - 1)
	] \big\}
\intertext{and shifting the sum}
   =&	\frac{1}{\Xi} \sum_{q=1}^\infty \sum_{k=q}^\infty \exp \big\{ -\beta [
		(1 - \mu)k + \lambda k (k - 1)
	] \big\}
\\
   =&	\frac{1}{\Xi} \sum_{k=1}^\infty \sum_{q=1}^k \exp \big\{ -\beta [
		(1 - \mu)k + \lambda k (k - 1)
	] \big\}
\\
   =&	\frac{1}{\Xi} \sum_{k=1}^\infty k \exp \big\{ -\beta [
		(1 - \mu)k + \lambda k (k - 1)
	] \big\}
\\
   =&	\frac{1}{\Xi} \trace_{\mathcal{F}(\CC)} \big[ n \exp \big\{ -\beta [
		(1 - \mu)n +\lambda n(n - 1)
	] \big\} \big]
\\
   =&	\rhodens.
\end{align*}
Therefore the absence of condensation implies that the sum of all finitely long cycle densities
equals the system density.
\hfill $\square$

\appendix

\renewcommand{\theequation}{A.\arabic{equation}}
\setcounter{equation}{0}  
\section*{Appendix A. Some useful inequalities and bounds}		\label{appendixA}

By the  operator inequalities $c_0^\ast c_0 \le N_\sV$, $\sum_x n_x^2 \ge \frac{N_\sV^2}{V}$ and
\[
	\nu \sqrt{V} (c\astr_0 + c^\ast_0 ) \le \nu^2 c^\ast_0 c\astr_0 + V \le \nu^2 N_\sV + V,
\]
it is clear that the Hamiltonian with sources, $\Hv{q}(\nu)$, is superstable for fixed 
$q\in\NN$ and $\lambda>0$, i.e.
\[
	\Hv{q}(\nu)
\ge	\lambda \frac{N_\sV^2}{V} - (\lambda + \mu + \nu^2) N_\sV - V.
\]

Using the Bogoliubov inequality (\ref{bogoliubov}), since $H_\sV(\beta,\mu-2q\lambda,\nu) 
- \Hv{q}(\beta,\mu,\nu) - q(q - \mu) = 2 q\lambda \sum_{i=2}^V n_i $, one may find that
\begin{equation}									\label{no-one-bound}
	\bigg\la \sum_{i=2}^V n_i \bigg\ra_{\Hv{q}(\beta,\mu,\nu)}
\le	\bigg\la \sum_{i=2}^V n_i \bigg\ra_{H_\sV(\beta,\mu-2q\lambda,\nu)}.
\end{equation}

Then we may find an upper bound (which is independent of $q$) for the expectation of the number 
operator with respect to $\Hv{q}(\nu)$, using (\ref{bog-5}) and (\ref{no-one-bound}), as follows:
\[	
	\big\la N_\sV \big\ra_{\Hv{q}(\nu)} 
= 	\big\la n_1 \big\ra_{\Hv{q}(\nu)} + \bigg\la \sum_{i=2}^V n_i \bigg\ra_{\Hv{q}(\nu)}
\le 	\bigg\la \frac{N_\sV}{V} \bigg\ra_{H_\sV(\nu)}
	+ \bigg\la \sum_{i=2}^V n_i \bigg\ra_{H_\sV(\beta, \mu-2q\lambda, \nu)}
\]
Now
\[
	\bigg\la \sum_{i=2}^V n_i \bigg\ra_{H_\sV(\beta, \mu-2q\lambda, \nu)}
\le	\big\la N_\sV \big\ra_{H_\sV(\beta, \mu-2q\lambda, \nu)}
\le	\big\la N_\sV \big\ra_{H_\sV(\beta, \mu, \nu)}
\]
since $\big\la N_\sV \big\ra_{H_\sV(\beta, \mu, \nu)}$ is monotonically increasing in $\mu$.
Hence
\begin{equation}									\label{bound-indepq}
	\big\la N_\sV \big\ra_{\Hv{q}(\nu)} 
\le 	\bigg\la \frac{N_\sV}{V} \bigg\ra_{H_\sV(\nu)} 
	+ \big\la N_\sV \big\ra_{H_\sV(\beta, \mu, \nu)}
\le	2\big\la N_\sV \big\ra_{H_\sV(\nu)}.
\end{equation}

By the superstability of the Hamiltonian $H_\sV(\nu)$, there exists a 
function $M \ge 0$ such that $\la N_\sV \ra_{H_\sV(\nu)} / V \le M$ for all $V$ and $\nu < \nu_0$. 
Using the fact that $\la a\astr_x \ra_{\Hv{q}(\nu)} = \la a^\ast_x \ra_{\Hv{q}(\nu)}$ 
and (\ref{bound-indepq}) we have
\begin{equation}										\label{bound-6}
	\frac{1}{V} \big\la c\astr_0 \big\ra_{\Hv{q}(\nu)}^2 
\le	\frac{1}{V} \big\la c_0^\ast c_0 \big\ra_{\Hv{q}(\nu)}
\le	\bigg\la \frac{N_\sV}{V} \bigg\ra_{\Hv{q}(\nu)} 		\!\!
\le	\frac{1}{2}\bigg\la \frac{N_\sV}{V} \bigg\ra_{H_\sV(\nu)} 		\!\!
\le	\frac{M}{2}
\end{equation}
for all $\nu < \nu_0$ and $q \in \NN$.

Considering the term $\left\la \delta_0^\ast \delta\astr_0 \right\ra_{\Hv{q}(\nu)}$ 
stated in Lemma \ref{lemma1},
we follow the procedure in Appendix 1 of Bru and Dorlas\cite{BruDorlas} to write
\begin{equation}										\label{expectation}
	\left\la \delta_0^\ast \delta\astr_0 \right\ra_{\Hv{q}(\nu)}
=	 (\delta_0, \delta_0 )_{\Hv{q}(\nu)} 
	+ \frac{\beta}{3} \left\la [ \delta^\ast_0, [ \Hv{q}(\nu) , \delta\astr_0]] \right\ra_{\Hv{q}(\nu)}
	- 1
\end{equation}
where the Duhamel inner product $( \, \cdot \, , \, \cdot \, )$ is defined as follows:
\begin{equation}										\label{bog-inner-prod}
	( A , B )_H = \frac{1}{\beta Z} \int_0^\beta \trace 
		\left[ A^\ast \e^{-(\beta - s) H} B \e^{-s H}\right] ds
\end{equation}
with $Z = \trace \e^{-\beta H}$. Since $\delta_0 = c_0 + \la c_0 \ra$ one may then evaluate that
\[
	[ \delta^\ast_0, [ \Hv{q}(\nu) , \delta\astr_0]]
=	[ c^\ast_0, [ \Hv{q}(\nu) ,c\astr_0]]
=	-\mu + \lambda + 2\lambda \frac{2N_\sV + q}{V}
\]
to give that (\ref{expectation}) has an upper bound of the form:
\begin{equation}										\label{bound-8}
	\left\la \delta_0^\ast \delta\astr_0 \right\ra_{\Hv{q}(\nu)}
\le	(\delta_0, \delta_0 )_{\Hv{q}(\nu)} +
	u \frac{q}{V} + w
\end{equation}
(using (\ref{bound-6})\!) for some constants $u$ and $w$ (independent of $q$).

We set $f_{q,\sV}(\nu) = (\delta_0, \delta_0 )_{\Hv{q}(\nu)}$. 
We wish to show that $f_{q,\sV}(\nu)$ is bounded by a constant independent of $q$.
For fixed $V$ and $\nu < \nu_0$, we have
\begin{equation}									\label{second-deriv-line1}
	4\beta f_{q,\sV}(\nu) 
=	\frac{1}{\nu} \frac{\partial}{\partial \nu} p_\sV[H_{q,\sV}(\nu)]
	+  \frac{\partial^2}{\partial \nu^2} p_\sV[H_{q,\sV}(\nu)].
\end{equation}

By (\ref{bound-6}), we have for all $q \in \NN$ that
\begin{equation}									\label{firstderiv-bound}
	\frac{\partial}{\partial \nu} p_\sV[\Hv{q}(\nu)] 
=	\frac{2}{\sqrt{V}} \left| \big\la c\astr_0 \big\ra_{\Hv{q}(\nu)} \right|
\le 	\sqrt{2M}.
\end{equation}
The second derivative term of (\ref{second-deriv-line1}) is:
\begin{align}	
	\beta^{-1} \frac{\partial^2}{\partial \nu^2} p_\sV[H_{q,\sV}(\nu)]
&=	\left( (c\astr_0 + c^\ast_0 )^\ast , c\astr_0 + c^\ast_0 \right)_{H_{q,\sV}(\nu)}
	- \big\la c\astr_0 + c^\ast_0 \big\ra_{H_{q,\sV}(\nu)}^2			\notag
\\
&\le	\left( c\astr_0 + c^\ast_0 , c\astr_0 + c^\ast_0 \right)_{H_{q,\sV}(\nu)}	\notag
\\
\intertext{using the fact that $(A,A) \le \tfrac{1}{2}\la A^\ast A + A A^\ast \ra$}
&\le	\big\la (c\astr_0 + c^\ast_0 )( c\astr_0 + c^\ast_0 ) \big\ra_{H_{q,\sV}(\nu)}	\notag
\\
&\le	V (4M+1 + 2\sqrt{2M(2M+1)} )						\label{f-bound-bad}
\end{align}
by the Schwarz inequality, (\ref{bound-indepq}) and (\ref{bound-6}).

\textbf{Acknowledgements:} The author would like to thank J.V. Pul\'e for his guidance, encouragement and 
many enlightening discussions, and the Irish Research Council for Science, Engineering and Technology
for their financial support.

%

\end{document}